\documentclass[12pt]{article}  
\pdfoutput=1
\usepackage{epsfig,epsf}  
\usepackage{graphics}  
\usepackage{cite}  
\usepackage{geometry}
\usepackage{xcolor} 
\usepackage{float}

\usepackage{amsmath}  
\usepackage{amsthm}  
\usepackage{amsfonts}  
\usepackage{amssymb} 
\usepackage{bm} 
\usepackage{mathtools}
\usepackage{slashed}
\usepackage{braket}

\usepackage{hyperref}
\usepackage{longtable}
\usepackage{caption}
\usepackage{feynmp-auto}
\usepackage{subcaption}
\usepackage{cleveref}

\numberwithin{equation}{section}

\newcommand{\Lbar}{\overline{\Lambda}}
\newcommand\gammat{\stackrel{\mathclap{\tiny\mbox{$(\sim)$}}}{\Gamma}}
\usepackage{booktabs}

\usepackage{amstext}
\usepackage{array}
\newcolumntype{C}{>{$}c<{$}}

\begin{document}
\allowdisplaybreaks
\thispagestyle{empty}  
  
\begin{flushright}  
{\small  
IPPP/21/38, SI-HEP-2021-026, SFB-257-P3H-21-071
\\[0.1cm]\today
}  
\end{flushright}  
  
\vskip1.5cm  
\begin{center}  
\textbf{\Large\boldmath $SU(3)$ breaking effects in $B$ and $D$ meson lifetimes}  
\\  
\vspace{1.5cm}  
{\sc Daniel~King}$^{(a)}$, {\sc Alexander~Lenz}$^{(b)}$ and {\sc Thomas~Rauh}$^{(c)}$\\[0.5cm]  
\vspace*{0.5cm} {\it 
$(a)$ IPPP, Department of Physics,
University of Durham,\\
DH1 3LE, United Kingdom\\[0.2cm]
$(b)$ CPPS,  Theoretische Physik 1, Department Physik,
Universität Siegen, Walter-Flex-Straße 3, 57068 Siegen,
Germany
\\[0.2cm]
$(c)$ Albert Einstein Center for Fundamental Physics,\\
Institute for Theoretical Physics, University of Bern,\\
Sidlerstrasse 5, CH-3012 Bern, Switzerland }
  
\def\thefootnote{\arabic{footnote}}  
\setcounter{footnote}{0}  
  
\vskip2.5cm  
\textbf{Abstract}\\  
\vspace{1\baselineskip}  


\parbox{0.9\textwidth}{
In the heavy quark expansion (HQE) of the total decay rates of $B_s$ and $D_s^+$ mesons non-perturbative
matrix elements of four quark operators are arising as phase space enhanced contributions.
We present the first determination of $m_s$ effects to the dimension six matrix
elements  of these four quark operators 
via a  heavy quark effective theory (HQET) sum rule analysis. 
In addition we calculate for the first time eye contractions of the four 
quark operators as well as matrix elements of penguin operators.
For the perturbative part we solve the 3-loop contribution to the sum  
rule and we evaluate condensate contributions.
In this study we work in the strict HQET limit and  our results can also be used to estimate
the size of the matrix element of the Darwin operator via equations of motion.
}
\end{center}  
  
\newpage
  
{\small \tableofcontents}  
  
\newpage  
\setcounter{page}{1} 


\section{Introduction}
The theoretical predictions for  $B$-meson lifetime ratios currently stand in close agreement with experimental results, see Table \ref{tab:my_label}.
\begin{table}[h]
    \centering
    \begin{tabular}{c|c|c}
    Lifetime Ratio & Experiment & Theory \\
    \hline 
    $\frac{\tau (B^+)}{\tau(B_d)}$  & $1.076 \pm 0.004$ \cite{Amhis:2019ckw}   & $1.078^{+0.021}_{-0.023}$\cite{Kirk:2017juj}\\
    \vspace{0.5mm} 
    $\frac{\tau (B_s)}{\tau(B_d)}$     & $0.998 \pm 0.005$ \cite{Amhis:2019ckw} & $1.0007\pm0.0025$ \cite{Kirk:2017juj} \\
\end{tabular}
    \caption{Experimental values (HFLAV \cite{Amhis:2019ckw}) of the lifetime ratio of $B$ mesons versus theoretical predictions based on the 2017 HQET sum rule prediction for the matrix elements of the four quark operators in the $\overline{MS}$  scheme \cite{Kirk:2017juj}.}
    \label{tab:my_label}
\end{table}
The measurements of the $B_s$ lifetime have recently been updated by
the LHCb collaboration \cite{Aaij:2019mhf,Aaij:2019vot},
the ATLAS collaboration \cite{Aad:2020jfw}
and by the CMS collaboration \cite{Sirunyan:2020vke} and
interestingly the value of ATLAS deviates from the other
measurements \cite{Lenz:2021bkv}.
In future we expect a further improvement of the experimental precision indicated in Table \ref{tab:my_label}.
On the theory side there has also been significant progress in the last years.
According to the heavy quark expansion (HQE)
\cite{Khoze:1983yp,Shifman:1984wx,Bigi:1991ir,Bigi:1992su,Blok:1992hw,Blok:1992he,Chay:1990da,Luke:1990eg,Bigi:1992ne}
(see Ref.~\cite{Lenz:2014jha} for a recent review) the total decay rate of a hadron $H_Q$ containing a heavy
quark $Q$ can be expanded in
inverse powers
 of the heavy quark mass $m_Q$ and each term in the expansion is a product of a perturbative coefficient $\Gamma_i$ or $\tilde \Gamma_i$
and a non-perturbative matrix element of a $\Delta Q = 0$ operator ${\cal O}_D$ or $\tilde{\cal O}_D$ of dimension $D$:
\begin{eqnarray}
\Gamma & = & \Gamma_3        \langle {\cal O}_{3} \rangle 
           + \Gamma_5  \frac{\langle {\cal O}_{5} \rangle}{m_Q^2} 
           + \Gamma_6  \frac{\langle {\cal O}_{6} \rangle}{m_Q^3} 
+ ...
+ 16 \pi^2 \left[
   \tilde{\Gamma}_6 \frac{\langle \tilde{{\cal O}}_{6} \rangle}{m_Q^3} 
+  \tilde{\Gamma}_7 \frac{\langle \tilde{{\cal O}}_{7} \rangle}{m_Q^4} 
+ ...
\right]\, ,
\label{eq:HQE}
\end{eqnarray}
with $\langle {\cal O}_{D} \rangle = \langle H_Q | {\cal O}_{D} | H_Q  \rangle/(2 M_{H_Q})$. 
We denote with $\Gamma_i$ contributions related to two quark operators ${\cal O}_i$
and with  $\tilde{\Gamma}_i$ contributions related to four quark  operators  $\tilde{\cal O}_i$. 
Each perturbative coefficient $\Gamma_i$ ($\tilde \Gamma_i$)
 can be further expanded in the strong coupling constant
\begin{equation}
  \gammat_i \, \, = \, \, {\gammat_i}^{(0)} \! + \,        \frac{\alpha_s}{4 \pi}  \,   \,      {\gammat_i}^{(1)}
  \! +  \left(\frac{\alpha_s}{4 \pi}\right)^2 \, {\gammat_i}^{(2)} + \dots \, .
\end{equation}
Traditionally the four quark contributions indicated by
$\tilde{\Gamma}_6 \langle \tilde{{\cal O}}_{6} \rangle$ are considered
to give the dominant contributions to lifetimes ratios, because
of the phase space enhancement factor $16 \pi^2$, see e.g.
Refs.~\cite{Uraltsev:1996ta,Neubert:1996we}. In these so-called spectator
contributions, which are known to NLO-QCD accuracy
\cite{Beneke:2002rj,Franco:2002fc,Ciuchini:2001vx,Keum:1998fd}, the by far largest source of uncertainty resides
in the non-perturbative  hadronic matrix elements $ \langle \tilde{{\cal O}}_{6} \rangle $. 
The most recent estimates for these parameters from lattice 
QCD \cite{Becirevic:2001fy} were carried out in 2001 and only made public in proceedings. 
In 2017 \cite{Kirk:2017juj} a significant improvement to the precision of the
dimension-6 matrix elements was achieved by means of a 3-loop HQET sum rule analysis.
In that case, spectator mass effects in the sum rule were neglected. This 
is a sensible simplification for $B^+$ and $B_d$ mesons, where the spectator
quark is an up or down quark. In the case of the $B_s$ meson
however, $SU(3)_F$ breaking effects are not expected to be negligible. 
In this paper we present the first computation of the dimension-6 matrix
elements of $\Delta Q = 0$ four quark operators with a non-zero strange quark mass, following the method established
in Ref.~\cite{King:2019lal}, where
$m_s$ effects to the HQET sum rules for $B_s$ mixing were calculated. These efforts lead to results with a competitive precision for $B$ mixing observables, see Ref.~\cite{DiLuzio:2019jyq}, as modern
lattice determinations \cite{FermilabLattice:2016ipl,Boyle:2018knm,Dowdall:2019bea}
and to strong bounds on BSM models that try to
explain the flavour anomalies, see e.g.
Refs.~\cite{DiLuzio:2019jyq,DiLuzio:2017fdq}.
In addition we determine for the first time eye contractions of 
the $\Delta Q = 0$ four quark operators as well as matrix elements of 
penguin operators.
\\
Very recently the Darwin term $\Gamma_6  \langle {\cal O}_{6} \rangle$ was
calculated for the first time for non-leptonic decays and found to be very
large \cite{Lenz:2020oce,Mannel:2020fts,MorenoTorres:2020xir}. For the
lifetime ratio $\tau (B^+) / \tau(B_d)$ this contribution will cancel due 
to isospin symmetry. However, for  a precise calculation of the ratio
$\tau (B_s) / \tau(B_d)$ the 
$SU(3)_F$ breaking contribution of the form $\Gamma_6  
(\langle {\cal O}_{6} \rangle_{B_d} - \langle {\cal O}_{6} \rangle_{B_s})$
has to be determined. 
The matrix element $\langle {\cal O}_{6} \rangle_{B_d} $ is known quite well from fits of the inclusive semileptonic $B$ meson decays, see 
e.g. Refs.~\cite{Alberti:2014yda,Bordone:2021oof}, unfortunately a corresponding analysis has not been performed for the $B_s$
meson, thus $\langle {\cal O}_{6} \rangle_{B_s}$
is largely unknown.
However, the Darwin operator can be related to four quark
operators via equations of motion (see e.g. \cite{Mannel:2020fts,King:2021xqp}) and thus our 
results can also be used to
estimate the size of the matrix element of the Darwin operator for the $B_s$ meson.
\\
In the following sections, we will restrict our discussion to the calculation
of the hadronic matrix elements themselves and reserve a full analysis of the
$B$ lifetimes for a subsequent paper in which the results presented here will
be used alongside other recent developments in the HQE
\cite{Lenz:2020oce,Mannel:2020fts,MorenoTorres:2020xir}.
\\
Since we work here in the strict HQET limit our results can also be applied to the charm sector, where sizeable 
lifetime differences have been found experimentally \cite{Zyla:2020zbs,Belle-II:2021cxx}:
\begin{equation}
    \frac{\tau (D^+)}{\tau (D_0)} = 2.54 \pm 0.02 \, ,
    \hspace{1cm}
    \frac{\tau (D_s^+)}{\tau (D_0)} = 1.20 \pm 0.01 \, .
    \end{equation}
As the expansion parameter $\alpha_s (m_c)$ and $\Lambda / m_c$ where
$\Lambda$ is a hadronic scale are quite sizeable, a study of charm lifetimes
can shed light on the convergence radius of the HQE \cite{King:2021xqp}.
\\
Our results can of course also be used for an analysis of spectator effects in inclusive semi-leptonic 
$B$ and $D$ meson decays, where the same matrix elements will appear, see 
e.g. Ref.~\cite{King:2021xqp}.
\\
 The rest of this paper is arranged as follows:
 Section \ref{sec:setup} consists of the sum rule
 setup and a collection of the analytic results.
 We introduce the operator basis and the
 parameterisation of the matrix elements in
 Section \ref{subsec:basis}, while Section
 \ref{subsec:thesumrule} is devoted
 to the presentation of the
 sum rule itself. The perturbative part of the sum rule is
 discussed in Section \ref{subsec:method} including
 a brief overview of the determination of $m_s$
 corrections as well as the introduction of the
 eye-contractions. Condensate contributions will be
 revisited in Section \ref{subsec:condensates} 
 and in Section \ref{subsec:analytic} we present
 analytic results. 
 In Section \ref{sec:numerics}  we summarise the
 findings of our numerical analysis, and in Section 
 \ref{sec:conclusions} we  conclude. 


\section{Setup and calculation}
\label{sec:setup}
\subsection{Operator Basis}
\label{subsec:basis}
We carry out the sum rule in the exact HQET limit in order to avoid mixing between operators of different
mass dimensions. 
The basis we use coincides with that  of Ref.~\cite{Franco:2002fc}, except for the naming of the colour-octett operators.
In the  HQET limit (denoted by the tilde)  we get 
\begin{eqnarray}
 \tilde{Q}_1^q & = & \bar{h}\gamma_\mu(1-\gamma^5)q
 \cdot 
 \bar{q}\gamma^\mu(1-\gamma^5)h,
 \hspace{1.0cm} 
 \tilde{T}_1^q = \bar{h}\gamma_\mu(1-\gamma^5)T^Aq
 \cdot
 \bar{q}\gamma^\mu(1-\gamma^5)T^Ah, \nonumber\\
 \tilde{Q}_2^q & = & \bar{h}(1-\gamma^5)q
 \cdot
 \bar{q}(1+\gamma^5)h,
 \hspace{1.85cm} \tilde{T}_2^q = \bar{h}(1-\gamma^5)T^Aq\cdot \bar{q}(1+\gamma^5)T^Ah,
 \label{eq:Lifetimes_HQET_operators}
\end{eqnarray}
where $h$ denotes the HQET field describing the heavy quark $Q$ with mass $m_Q$, the light quark fields are denoted by $q$. In addition we use the same  evanescent operators as in Ref.~\cite{Kirk:2017juj}  (choosing $a_1=a_2=-8$). 
A full description of SU(3) flavour-breaking contributions at NLO in QCD also requires us to consider the QCD penguin operators 
\begin{equation}
 \tilde{Q}_P^q = \bar{h}\gamma_\mu T^A h \cdot \bar{q}\gamma^\mu T^A q\,.
 \label{eq:Penguin_operator}
\end{equation}
Note, that differing from the definition in Ref.~\cite{Franco:2002fc} we need the flavour specific contribution
of the penguins, thus we are not summing over the light quark flavour $q$.
Inspired by Refs.~\cite{Lenz:2013aua,Franco:2002fc} we parametrize the 
matrix elements of the above operators as,
\begin{align}
 \braket{{\mathbf{B}}_q|\tilde{Q}_i^q(\mu)|{\mathbf{B}}_q} & = A_{\tilde{Q}_i}F_q^2(\mu)\tilde{B}_i^q(\mu) \, & \braket{{\mathbf{B}}_q|\tilde{Q}_i^{q'}(\mu)|{\mathbf{B}}_q} & = A_{\tilde{Q}_i}F_q^2(\mu)\tilde{\delta}_i^{q'q}(\mu)\, ,
 \nonumber\\
 \braket{{\mathbf{B}}_q|\tilde{T}_i^q(\mu)|{\mathbf{B}}_q} & = A_{\tilde{T}_i}F_q^2(\mu)\tilde{\epsilon}_i^q(\mu) & \braket{{\mathbf{B}}_q|\tilde{T}_i^{q'}(\mu)|{\mathbf{B}}_q} & = A_{\tilde{T}_i}F_q^2(\mu)\tilde{\delta}_{i+2}^{q'q}(\mu)\, ,
 \nonumber\\
 \braket{{\mathbf{B}}_q|\tilde{Q}_P^q(\mu)|{\mathbf{B}}_q} & =
  A_{\tilde{Q}_P}F_q^2(\mu)\tilde{B}_P^q(\mu) & \braket{{\mathbf{B}}_q|\tilde{Q}_P^{q'}(\mu)|{\mathbf{B}}_q} & = A_{\tilde{Q}_P}F_q^2(\mu)\tilde{\delta}_{P}^{q'q}(\mu)\, ,
\label{eq:Matrix_Elements}
\end{align}
for which the colour factors correspond to,
\begin{equation}
  A_{\tilde{Q}_i}=A_{\tilde{T}_i}=1 \hspace{1cm}
  A_{\tilde{Q}_P}=-\frac{C_F}{2N_c}\,,
\label{eq:ColourFactors}
\end{equation}
with the HQET decay constant $F_q$, the bag parameters $ \tilde{B}_i^q$, $\tilde{B}_P^q$ and $ \tilde{\epsilon}_i^q$  and the non-valence contribution $\tilde{\delta}_i^{q'q} $, for $q \neq q'$.
Note that differing from Refs.~\cite{Franco:2002fc} and \cite{Lenz:2013aua} we have included
in $ \tilde{B}_i^q$, $\tilde{B}_P^q$ and $ \tilde{\epsilon}_i^q$ also the non-valence contributions with
 $q = q'$. As usual $\mu$ denotes the renormalisation scale dependence.
 In addition the heavy $| {\mathbf{B}}_q \rangle$ meson states 
 (consisting of a heavy anti-quark $\overline{Q}$ and a light quark ${q}$ )
 are  considered in the strict HQET limit
 and thus our expressions hold both for $B$ and $D$ mesons.

\subsection{The Sum Rule}
\label{subsec:thesumrule}
The HQET Borel sum rule for the decay constant $F_q$, as derived in
Refs.~\cite{Neubert:1991sp,Bagan:1991sg,Broadhurst:1991fc,Shuryak:1981fza}, 
is well studied. The starting point for its derivation is the 2-point correlator,
\begin{equation}
    \Pi(\omega)=\int \; d^d x e^{i p\cdot x}\Braket{0|\text{T}\left\{\tilde{j}_q(0)\tilde{j}_q^\dagger(x)\right\}|0}
\label{eq:2pointCorrelator}
\end{equation}
for a  heavy meson 
with the momentum $p_M = m_Q v + p$, where $v$ is the four-velocity of the meson and $p$ the residual momentum.
The  residual energy is denoted by  $\omega=p\cdot v$. 
The interpolating heavy meson current used in Eq.(\ref{eq:2pointCorrelator}) is defined as,
\begin{equation}
 \tilde{j}_q = \bar{q}\gamma^5 h.
\end{equation} 
The sum rule for $F_q$ then takes the form of,
\begin{equation}
    F^2_q(\mu)=\int\limits_0^{\omega_c}d\omega \; e^{\frac{\overline{\Lambda}_q-\omega}{t}}\rho_{\Pi}(\omega)
\label{eq:DecayConstantSR}
\end{equation}
for which $\rho_{\Pi}(\omega)$ is defined as the discontinuity of Eq.(\ref{eq:2pointCorrelator}), $\overline{\Lambda}_q$ is the meson mass-difference, the Borel parameter $t$ determines the degree to which continuum states of the hadronic spectral function are exponentially suppressed, and where we have introduced a cutoff of $\omega_c$.

In order to build a sum rule for the bag parameters however, the central object of the calculation is the 3-point correlator,
\begin{equation}
 K_{\tilde{\mathcal{O}}^{q'}}^q(\omega_1,\omega_2) = \int d^dx_1 d^dx_2 e^{i(p_1\cdot x_1- p_2\cdot x_2)}
 \braket{0|\text{T}\left\{\tilde{j}_q(x_2)\tilde{\mathcal{O}}^{q'}(0)\tilde{j}_q^\dagger(x_1)\right\}|0},
 \label{eq:DefK}
\end{equation}
where $p_i$ corresponds to the residual momentum of the incoming and outgoing states respectively, each with velocity $v$, and residual energy $\omega_{1,2} = p_{1,2}\cdot v$, and in addition to the heavy quark currents there is now also the insertion of a four quark operator.

\begin{figure}
\begin{subfigure}{.495\textwidth}
  \centering
\scalebox{0.65}{\includegraphics[width=\textwidth]{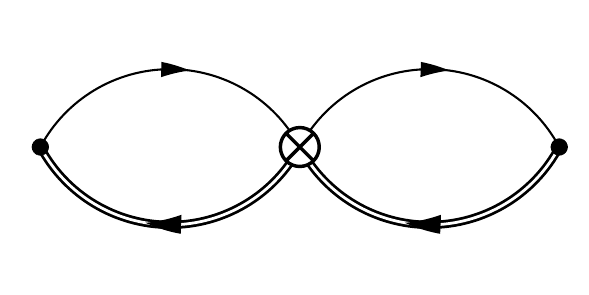}
}
  \caption{LO Factorisable}
  \label{fig:FactLO}
\end{subfigure}
\begin{subfigure}{.495\textwidth}
  \centering
\scalebox{0.65}{\includegraphics[width=\textwidth]{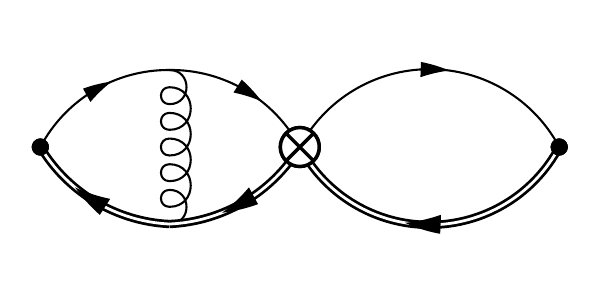}
}
  \caption{NLO Factorisable}
  \label{fig:FactNLO}
\end{subfigure}
\begin{subfigure}{.495\textwidth}
  \centering
\scalebox{0.65}{\includegraphics[width=\textwidth]{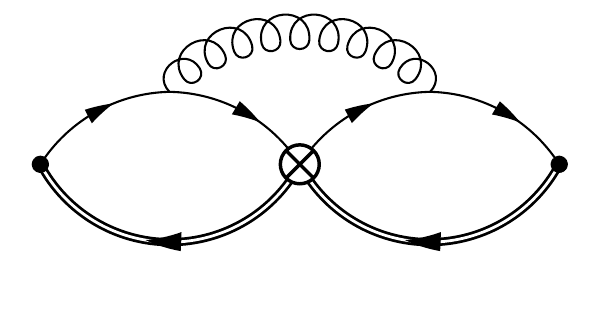}
}
  \caption{NLO Nonfactorisable}
  \label{fig:NonFactNLO}
\end{subfigure}
\begin{subfigure}{.495\textwidth}
  \centering
\scalebox{0.65}{\includegraphics[width=\textwidth]{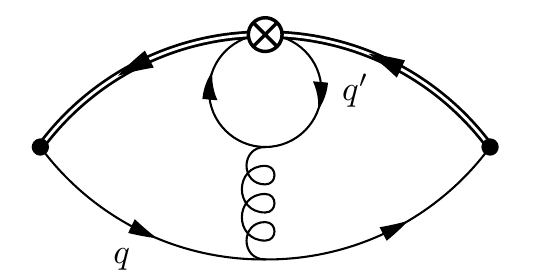}
}
  \caption{NLO Non-Valence}
  \label{fig:NonValNLO}
\end{subfigure}
\caption{Some of the diagrams contributing to the correletor in Eq.(\ref{eq:DefK}). 
Non-valence type diagrams like that shown in (d) only appear at NLO.}
\label{fig:3loops}
\end{figure}

As in Refs.~\cite{Kirk:2017juj,King:2019lal} we categorise the possible field contractions of Eq.(\ref{eq:DefK}) into
factorisable and non-factorisable contributions. Examples of the corresponding Feynman diagrams are found in
Fig.\ref{fig:3loops}. This separation of contributions allows us to formulate a sum rule for the deviation of the bag parameter 
$\Delta B$ from its vacuum saturation approximation (VSA) value. 
\begin{eqnarray}
\tilde{B}_i^q(\mu) & = &  1 + \Delta B^q_{\tilde{Q}_i^q}(\mu) \, ,
\\
 \tilde{\epsilon}_i^q(\mu)  & = &  0 + \Delta B^q_{\tilde{T_i^q}}(\mu) \, ,
\\
\tilde{B}_P^q(\mu)   & = &  1 + \Delta B^q_{\tilde{Q}_P^q}(\mu) \, .
\end{eqnarray}
We find the following  finite energy Borel sum rules,
\begin{equation}
    \Delta B^q_{\tilde{\mathcal{O}}^q}(\mu)=\frac{1}{A_{\mathcal{O}}F_q^4(\mu)}\int\limits_{0}^{\omega_c}d\omega_1 d\omega_2 e^{\frac{\overline{\Lambda}_q-\omega_1}{t}+\frac{\overline{\Lambda}_q-\omega_2}{t}} \Delta \rho_{\tilde{\mathcal{O}}^q}^q(\omega_1,\omega_2)
\label{eq:SumRuleVal}
\end{equation}
in which the term $\Delta \rho_{\tilde{\mathcal{O}}^q}^q(\omega_1,\omega_2)$ corresponds to the non-factorisable part of the double discontinuity of Eq.(\ref{eq:DefK}). Looking at the  whole double discontinuity of the 3-point correlator it is useful to separate out the various contributions further as,
\begin{equation}
\begin{split}
    \rho_{\tilde{\mathcal{O}}^q}^q(\omega_1,\omega_2) &= \delta_{\tilde{\mathcal{O}}\tilde{
    Q}}\rho_{\Pi}(\omega_1)\rho_{\Pi}(\omega_2)+
    \Delta \rho_{\tilde{\mathcal{O}}^q}^q(\omega_1,\omega_2) \\
    &= \delta_{\tilde{\mathcal{O}}\tilde{
    Q}}\rho_{\Pi}(\omega_1)\rho_{\Pi}(\omega_2)+
    \Delta_{\text{tree}} \rho_{\tilde{\mathcal{O}}^q}^q(\omega_1,\omega_2)+
    \Delta_{\text{peng}} \rho_{\tilde{\mathcal{O}}^q}^q(\omega_1,\omega_2) \;,
    \end{split}
\label{eq:doublediscontinuity}
\end{equation}

for which $\delta_{\tilde{\mathcal{O}}\tilde{Q}}$ is equal to 1 for the colour singlet and penguin operators and 0 for the colour octet operators. 
In Eq.(\ref{eq:doublediscontinuity}), the first term corresponds to factorisable contributions, whilst  $\Delta_{\text{tree}}\rho$ corresponds to the first set of non-factorisbale contractions (see Fig.\ref{fig:NonFactNLO}). The term denoted by $\Delta_{\text{peng}}\rho$, stems from `eye-contraction' diagrams like that of the example
illustrated in Fig.\ref{fig:NonValNLO} and was not considered in \cite{Kirk:2017juj} as
these diagrams only become necessary when taking into account SU(3) flavour breaking effects. Note also, that these
contributions first appear at NLO in QCD. The presence of the non-valence terms also forces us to expand our
basis of operators to include the penguin operator defined in Eq.(\ref{eq:Penguin_operator}), which arises in
renormalisation and thus mixes with the original basis under renormalisation group (RG) running. Details of the correlator renormalisation and the resulting structure of the renormalisation group equations (RGE) are presented in Appendix\,\ref{sec:RGE}.  For the matrix elements of operators with a
different light-quark flavour to that of the external meson state, only these `eye-contraction' diagrams
contribute and so the sum-rule has the form,
\begin{equation}
    \delta^{q'q}_{\tilde{\mathcal{O}}^{q'}}(\mu)=\frac{1}{A_{\mathcal{O}}F_q^4(\mu)}\int\limits_{0}^{\omega_c}{d\omega_1 d\omega_2 e^{\frac{\overline{\Lambda}_q-\omega_1}{t}+\frac{\overline{\Lambda}_q-\omega_2}{t}} \Delta_{\text{peng}} \rho_{\tilde{\mathcal{O}}^{q'}}^q(\omega_1,\omega_2)}
\label{eq:SumRuleNonVal}
\end{equation}
We further split the non-factorisables part into perturbative and condensate contributions 
\begin{equation}
    \Delta_{\text{tree/peng}} \rho_{\tilde{\mathcal{O}}^{q'}}^q(\omega_1,\omega_2) = 
    \Delta_{\text{tree/peng}}^\text{pert} \rho_{\tilde{\mathcal{O}}^{q'}}^q(\omega_1,\omega_2) + 
    \Delta_{\text{tree/peng}}^\text{cond} \rho_{\tilde{\mathcal{O}}^{q'}}^q(\omega_1,\omega_2)\,,
\end{equation}
where the tree contribution vanishes if $q\neq q'$. 
Those will be discussed in greater depth in Section \ref{subsec:method} and \ref{subsec:condensates} and we show the results of our calculations in Section~\ref{subsec:analytic}. 
\subsection{Perturbative contributions}
\label{subsec:method} 
As already mentioned, the eye-contraction diagrams represent a new contribution, not
previously calculated, to our sum rule analysis of the $B_{s}$ lifetime matrix elements.
The procedure for computing them however, is unchanged from that of the standard
tree-contraction terms, which was performed and  described in detail in
Ref.~\cite{Kirk:2017juj,King:2019lal}. So here we will only briefly summarise
our approach. 
\\
The three-point correlators were calculated using two separate implementations. In one, the amplitudes of the 3-loop processes were generated manually and the Dirac algebra was computed using Tracer \cite{Jamin:1991dp},
treating $\gamma^5$ in accordance with the Larin scheme \cite{Larin:1993tq}. Alternatively, the amplitudes were generated using QGRAF \cite{Nogueira:1991ex} and the Dirac algebra computed using a private implementation in the `NDR' scheme. We found full agreement between both computations. An `Integration by Parts' \cite{Chetyrkin:1981qh} reduction of the amplitudes was
carried out using FIRE5 \cite{Smirnov:2014hma}. The resulting master integrals are already
known \cite{Grozin:2008nu} to all orders in $\epsilon$ and we expanded these to the
required order using the HypExp package \cite{Huber:2007dx}. These master integrals
describe the massless spectator quark scenario. In order to compute the $m_s$ corrections,
an `expansion by regions' approach was taken \cite{Beneke:1997zp,Jantzen:2011nz}
and leads to a Taylor expansion in $m_s/\omega$, allowing us to `recycle' the master
integrals of the massless case. 
\\
In this work, it became apparent that when considering $m_s$ corrections to 
the eye-contractions, these contributions could only be treated consistently within the
traditional sum rule approach and not with the weight function method used in
Refs.~\cite{Kirk:2017juj,King:2019lal}. Therefore in this work we explicitly 
evaluate the
integrals in Eq.(\ref{eq:SumRuleVal}) and Eq.(\ref{eq:SumRuleNonVal}) in the traditional
sum rule framework. However, when applicable we compare the results of both methods and
show that we find them to be consistent. This will be discussed further in Sections
\ref{subsec:analytic} and \ref{sec:numerics}. 
There are two major consequences resulting from shifting towards a traditional sum rule approach. The
first is that when also using the HQET sum rule result for the decay constant (as shown in
Eq.(\ref{eq:DecayConstantSR})), the dependency of the bag parameter on
$\overline{\Lambda}_q$ drops out. The second is that there is now an explicit dependence
of the bag parameter on both the cut-off $\omega_c$ and the Borel parameter $t$. In our
implementation, these parameters were set by fixing the HQET sum rules for $F_q$ and
$\overline{\Lambda}_q$ to values found in the literature. Details of this procedure can be found in Appendix\,\ref{sec:FandLambda_SR}.\\

\subsection{Condensate contributions}
\label{subsec:condensates} 
\begin{figure}
\begin{subfigure}{.495\textwidth}
  \centering
\scalebox{0.9}{\includegraphics[width=\textwidth]{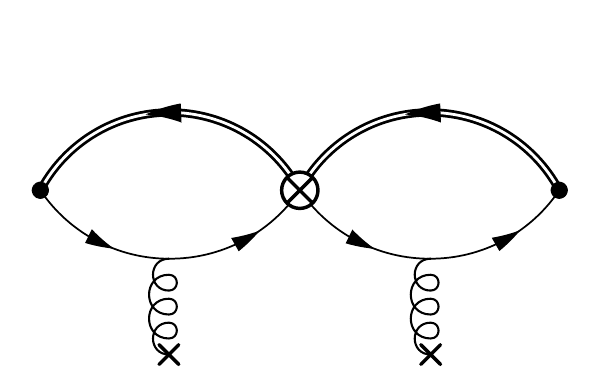}
}
  \label{fig:GG}
\end{subfigure}
\begin{subfigure}{.495\textwidth}
  \centering
\scalebox{0.9}{\includegraphics[width=\textwidth]{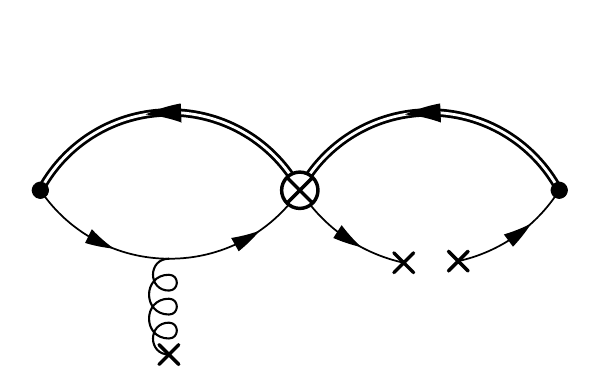}
}
  \label{fig:qGq}
\end{subfigure}
\caption{Condensate corrections corresponding to $\braket{\frac{\alpha_s}{4\pi}GG}$ and $\braket{g_s \overline{q}\sigma_{\mu\nu}G^{\mu\nu}q}$ respectively.}
\label{fig:Condensates_Non-vanishing}
\end{figure}

\begin{figure}
\centering
\begin{subfigure}{.495\textwidth}
    \centering
    \scalebox{0.85}{\includegraphics[width=\textwidth]{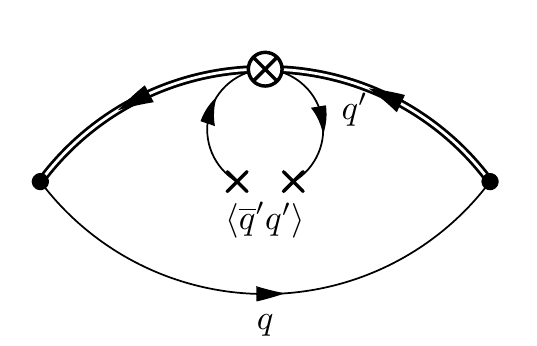}}
    \label{fig:q'q'}
\end{subfigure}
\begin{subfigure}{.495\textwidth}
    \centering
    \scalebox{0.85}{\includegraphics[width=\textwidth]{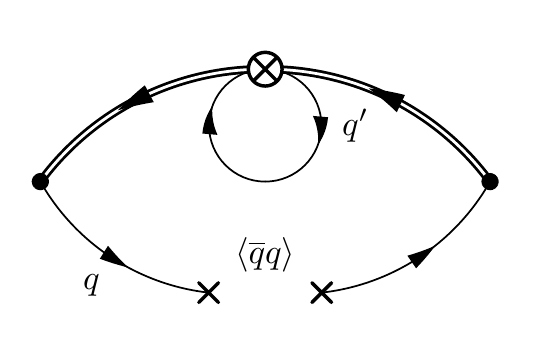}}
    \label{fig:qq}
\end{subfigure}
\caption{Quark condensate contributions to the eye contractions at leading order in $\alpha_s$.}
\label{fig:ThreePointCorrelatorEyeDim3}
\end{figure}

We also carried out an independent analysis of the condensate contributions, that have
previously been determined for the massless case in
Refs.~\cite{Cheng:1998ia,Baek:1998vk}\footnote{In the paper by Baek 
\textit{et al.} \cite{Baek:1998vk}, Eq.(20) yields an additional factor of 4
for $\epsilon_1$ compared to the expression found in Eq.(11) of the same
paper.} for the $Q_i$ and $T_i$ operators, but not for $Q_P$.
Whenever appropriate we compare our results with the literature in  Section \ref{subsec:analytic}.
\\
We use the  standard approach of the background field method
\cite{Novikov:1983gd,Reinders:1984sr,Pascual:1984zb}. Since, in calculating the deviation
of the bag parameters from their VSA values, we are only concerned with non-factorisable
contributions, the only diagrams that need to be considered are those found in
Fig.~\ref{fig:Condensates_Non-vanishing} along with their symmetric counterparts. These
represent the only condensate corrections up to dimension-6 and leading order in
$\alpha_s$, assuming that the quark condensate factorises and thus leads to no correction to the non-factorisable contribution.
\\
With regards to the non-valence terms, there is no dimension three 
quark condensate contribution at the leading order in $\alpha_s$ from the diagrams in 
Fig.~\ref{fig:ThreePointCorrelatorEyeDim3}. The left diagram vanishes because the quark
condensate  flips the chirality and the Dirac structure
$\Gamma_1\langle\bar{q}'q'\rangle\Gamma_2$ vanishes for all the 
combinations of currents $\Gamma_{1,2}$ appearing in the considered operators. The
$\langle\bar{q}'q'\rangle$  condensate is therefore suppressed by an additional $\alpha_s
m_{q'}/\Lbar$. The right diagram is scaleless.  The $\langle\bar{q}q\rangle$ condensate is
therefore suppressed by at least an extra $\alpha_s$.  There is also no dimension four
gluon condensate $\langle\alpha_s G^2\rangle$ contribution at leading order 
in $\alpha_s$ because the penguin loop is scaleless without an extra gluon. Similar
arguments lead us to  conclude that the dimension five quark gluon condensate
$\langle\bar{q}^{(')}\sigma_{\mu\nu}G^{\mu\nu}q^{(')}\rangle$  and the dimension six quark
condensate $\langle\bar{q}'q'\bar{q}q\rangle$\footnote{If $q'=q$ this does not vanish, but
is part of the factorisable contribution.} do not contribute at leading order  in
$\alpha_s$. Therefore, condensate contributions to the eye-contractions are
suppressed with respect to the perturbative contribution at first order in the
strong coupling and are not taken into account. 

\subsection{Analytic results}
\label{subsec:analytic} 
In this section we present the analytic expressions of our calculation. Beginning with the perturbative contribution, the double discontinuities defined in Eq.(\ref{eq:doublediscontinuity}) can be expressed in terms of
their $m_s$ (generally denoted as $m_q$ below) expansion as, 
\begin{align}
 \Delta_\text{tree}^\text{pert}\rho_{\tilde{\mathcal{O}}^{q}}^q(\omega_1,\omega_2) \equiv \,
 & \frac{N_c C_F}{4} \frac{\omega_1^2\omega_2^2}{\pi^4}\frac{\alpha_s}{4\pi}
 \Bigg[r_{\tilde{\mathcal{O}}}^{(0)}(x,L_\omega) + \left(\frac{m_q}{\omega_1} + \frac{m_q}{\omega_2}\right)r_{\tilde{\mathcal{O}}}^{(1)}(x,L_\omega) \nonumber
 \\
 & + \left(\frac{m_q^2}{\omega_1^2} + \frac{m_q^2}{\omega_2^2}\right)r_{\tilde{\mathcal{O}}}^{(2)}(x,L_\omega) 
 + \dots \Bigg]\,\theta(\omega_1-m_q)\theta(\omega_2-m_q),
 \label{eq:r_def}
 \end{align}
for $x=\omega_2/\omega_1$ and $L_{\omega}=\ln(\mu^2/(4\omega_1 \omega_2))$. 
 \\
The non-factorisable tree contributions for the colour singlet operators at order $\alpha_s$ 
have a vanishing color factor, yielding $r_{\tilde{Q}_i}^{(j)} = 0$. In the massless 
limit we find~\cite{Kirk:2017juj} 
\begin{eqnarray}
 r_{\tilde{T}_1}^{(0)} & = & -8+\frac{a_1}{8}+\frac{2 \pi ^2}{3}-\frac{3}{2}L_\omega-\frac{1}{4}\phi(x)\,, \nonumber\\
 r_{\tilde{T}_2}^{(0)} & = & -\frac{29}{4}+\frac{a_2}{8}+\frac{2 \pi ^2}{3}-\frac{3}{2}L_\omega-\frac{1}{4}\phi(x)\,,
 \label{eq:Lifetimes_r0}
\end{eqnarray}
and 
\begin{equation}
r_{\tilde{Q}_P}^{(0)} = \frac{1}{8N_c}\left[-30+\frac{8 \pi ^2}{3}-6L_\omega-\phi(x)\right]\,, 
\end{equation}
for the penguin operator, where 
\begin{equation}
 \phi(x)=\begin{cases}
  x^2-8 x+6 \ln(x),\hspace{1cm}x\leq1,\\ 
  \frac{1}{x^2}-\frac{8}{x}-6 \ln(x),\hspace{1.2cm}x>1.
 \end{cases}
\end{equation}
The linear terms in the strange quark mass read 
\begin{eqnarray}
 r_{\tilde{T}_1}^{(1)} & = & \frac{a_1}{8}+\frac{2 \pi ^2}{3}-\frac{3}{2}L_\omega - \begin{cases}
       \frac{2 (36+9 x+x^2)}{9 (1+x)}+\frac{9+9 x-2 x^2}{6 (1+x)} \ln (x), & x\leq1,\\
       \frac{2(1+9 x+36 x^2)}{9 x (1+x)}+\frac{2-9 x-9 x^2}{6 x (1+x)}\ln (x), & x>1,
       \end{cases}\nonumber\\
 r_{\tilde{T}_2}^{(1)} & = & \frac{a_2}{8}+\frac{2 \pi ^2}{3}-\frac{3}{2}L_\omega + \begin{cases}
       -\frac{29+11 x-2 x^2}{4 (1+x)}-\frac{3}{2}\ln (x), & x\leq1,\\
       \frac{2-11 x-29 x^2}{4 x (1+x)}+\frac{3}{2}\ln (x), & x>1,
       \end{cases}\\
 r_{\tilde{Q}_P}^{(1)} & = & \frac{1}{36N_c}\Bigg[-12 \pi ^2+27L_\omega + \begin{cases}
       \frac{135+81 x-22 x^2}{1+x}+\frac{3(9+9x+2x^2)}{1+x}\ln (x), & x\leq1,\\
       -\frac{22-81 x-135 x^2}{x (1+x)}-\frac{3(2+9x+9x^2)}{x(1+x)}\ln (x), & x>1,
       \end{cases} \Bigg]\nonumber
 \label{eq:Lifetimes_r1}
\end{eqnarray}
and for the corrections quadratic in $m_s$ we find 
\begin{eqnarray}
 r_{\tilde{T}_1}^{(2)} & = & \frac{1}{1+x^2} \left[-\frac{(1-x)^2a_1}{16}+\frac{3(1-x)^2}{4} L_\omega -\frac{x}{4}\psi(x) \left(1+\frac{3(1+x)}{1-x}\ln(x)\right) \right.\nonumber\\
 \nonumber\\
   & & \left.+ \begin{cases}
       \frac{\pi^2(1+8 x-5 x^2)}{12} + \frac{24-48 x+16 x^2+x^3}{6} + \frac{1+x^2}{2}\ln(x) &\\
       + \frac{1-x^2}{2}\ln^2(x) + \frac{5(1-x^2)}{2}\text{Li}_2\left(1-\frac{1}{x}\right), & x\leq1,\\
       \\
       \frac{\pi^2(-5+8 x+x^2)}{12} + \frac{1+16 x-48x^2+24x^3}{6x} - \frac{1+x^2}{2}\ln(x) &\\
       - \frac{1-x^2}{2}\ln^2(x) - \frac{5(1-x^2)}{2}\text{Li}_2(1-x), & x>1,
       \end{cases}\right],\nonumber\\
  \nonumber\\      
 r_{\tilde{T}_2}^{(2)} & = & \frac{1}{1+x^2} \left[-\frac{a_2(1-x)^2}{16} - \frac{\pi^2(1-4 x+x^2)}{6} + \frac{3(1-x)^2}{4}L_\omega \right.\nonumber\\
   & & + \frac{29-62 x+29x^2}{8} -\frac{x}{2}\psi(x)\left(1+\frac{1+x}{1-x}\ln(x)\right) \nonumber\\
   & & \left. + \begin{cases}
       \frac{(1-x)^2}{4}\ln(x) + (1-x^2)\text{Li}_2\left(1-\frac{1}{x}\right), & x\leq1,\\
       -\frac{(1-x)^2}{4}\ln(x) - (1-x^2)\text{Li}_2\left(1-x\right), & x>1,
       \end{cases} \right],\\
 \nonumber\\ 
 r_{\tilde{Q}_P}^{(2)} & = & \frac{1}{24N_c(1+x^2)} \left[9(1-x)^2 L_\omega - 9x\psi(x)\left(1+\frac{1+x}{3(1-x)}\ln(x)\right)\right.\nonumber\\
   & & \left. + \begin{cases}
       45-102x+61x^2-2x^3 - (5-8x-x^2)\pi^2 - 12x\ln(x) &\\
       - 6(1-x^2)\ln^2(x) - 6(1-x^2)\text{Li}_2\left(1-\frac{1}{x}\right), & x\leq1,\\
       \\
       -\frac{2-61x+102x-45x^3}{x} + (1+8x-5x^2)\pi^2 + 12x\ln(x) &\\
       + 6(1-x^2)\ln^2(x) + 6(1-x^2)\text{Li}_2\left(1-x\right), & x>1,
       \end{cases} \right]\nonumber
 \label{eq:Lifetimes_r2}
\end{eqnarray}
with 
\begin{equation}
 \psi(x) = \begin{cases}
            \frac{(1-x)^2}{x}\left[2\ln(1-x) - \ln(x)\right], \hspace{1cm}x\leq1,\\
            \frac{(1-x)^2}{x}\left[2\ln(x-1) - \ln(x)\right], \hspace{1cm}x>1.
           \end{cases}
\end{equation}
The perturbative contribution to the double discontinuities
of the eye-contractions, defined in Eq.(\ref{eq:doublediscontinuity}), can be expressed in terms of
their $m_s$ (generally denoted as $m_q$ and $m_{q'}$ below) expansion as
\begin{align}
 \Delta_\text{peng}^\text{pert}\rho_{\tilde{\mathcal{O}}^{q'}}^q(\omega_1,\omega_2) \equiv \,& \frac{N_c C_F}{4} \frac{\omega_1^2\omega_2^2}{\pi^4}\frac{\alpha_s}{4\pi}
 \Bigg[s_{\tilde{\mathcal{O}}}^{(0)}(x,L_\omega) + \left(\frac{m_q}{\omega_1} + \frac{m_q}{\omega_2}\right)s_{\tilde{\mathcal{O}}}^{(1)}(x,L_\omega) \nonumber\\
 & + \left(\frac{1}{\omega_1^2} + \frac{1}{\omega_2^2}\right)\left[m_q^2s_{\tilde{\mathcal{O}}}^{(2)}(x,L_\omega) + m_{q'}^2t_{\tilde{\mathcal{O}}}^{(2)}(x,L_\omega)\right] 
 + \dots \Bigg]\nonumber\\
 & \times\theta(\omega_1-m_q)\theta(\omega_2-m_q).
\label{eq:st_def}
\end{align}
For the non-valence expression Eq.(\ref{eq:st_def}), $s_{\tilde{\mathcal{O}}}^{(i)}$
corresponds to $m_s$ corrections of order $i$ stemming from a non-zero $q$ quark mass (see
Fig.\ref{fig:3loops}), whereas $m_s$ corrections attributed to the $q'$ quark are contained
within the $t_{\tilde{\mathcal{O}}}^{(2)}$ term. It is also worth noting that there is no
$t_{\tilde{\mathcal{O}}}^{(1)}$ in Eq.(\ref{eq:st_def}) since the double discontinuity
evaluates to zero.

At the considered order the eye contributions for the color singlet and octet operators differ 
only by their color factors 
\begin{equation}
 s_{\tilde{T}_i}^{(j)} = \frac{-1}{2N_c}\,s_{\tilde{Q}_i}^{(j)},\hspace{1cm}t_{\tilde{T}_i}^{(2)} = \frac{-1}{2N_c}\,t_{\tilde{Q}_i}^{(2)}. 
\end{equation}
Our results for the singlet and penguin operators are in the massless case
\begin{eqnarray}
 s_{\tilde{Q}_1}^{(0)} & = & \frac{20}{9}+\frac{2}{3}L_\omega+\frac{1}{9}\phi(x), \nonumber\\
 s_{\tilde{Q}_2}^{(0)} & = & -\frac{13}{9}-\frac{1}{3}L_\omega-\frac{1}{18}\phi(x), \nonumber\\
 s_{\tilde{Q}_P}^{(0)} & = & \frac{13}{9}+\frac{1}{3}L_\omega+\frac{1}{18}\phi(x).
 \label{eq:Lifetimes_s0}
\end{eqnarray}
The corrections proportional to the strange quark mass read
\begin{eqnarray}
 s_{\tilde{Q}_1}^{(1)} & = & \frac{2}{3}L_\omega + \begin{cases}
       \frac{2 (10+x-x^2)}{9 (1+x)}+\frac{2}{3} \ln (x), & x\leq1,\\
       -\frac{2(1-x-10 x^2)}{9 x (1+x)}-\frac{2}{3}\ln (x), & x>1,
       \end{cases},\nonumber\\
 s_{\tilde{Q}_2}^{(1)} & = & -\frac{1}{3}L_\omega + \begin{cases}
       -\frac{13+4 x-x^2}{9 (1+x)}-\frac{1}{3}\ln (x), & x\leq1,\\
       \frac{1-4 x-13 x^2}{9 x (1+x)}+\frac{1}{3}\ln (x), & x>1,
       \end{cases},\nonumber\\
 s_{\tilde{Q}_P}^{(1)} & = & -\frac{1}{3}L_\omega + \begin{cases}
       -\frac{13+4 x-x^2}{9 (1+x)}-\frac{1}{3}\ln (x), & x\leq1,\\
       \frac{1-4 x-13 x^2}{9 x (1+x)}+\frac{1}{3}\ln (x), & x>1,
       \end{cases}  
 \label{eq:Lifetimes_s1}
\end{eqnarray}
while the corrections quadratic in $m_s$ are given by
\begin{eqnarray}
 s_{\tilde{Q}_1}^{(2)} & = & \frac{1}{1+x^2}\left[-\frac{10(1-x)^2}{9}-\frac{(1-x)^2}{3}L_\omega+\frac{x}{3}\psi(x)\right], \nonumber\\
 s_{\tilde{Q}_2}^{(2)} & = & \frac{1}{1+x^2}\left[\frac{13(1-x)^2}{18}+\frac{(1-x)^2}{6}L_\omega-\frac{x}{6}\psi(x)\right], \nonumber\\
 s_{\tilde{Q}_P}^{(2)} & = & \frac{1}{1+x^2}\left[-\frac{13(1-x)^2}{18}-\frac{(1-x)^2}{6}L_\omega+\frac{x}{6}\psi(x)\right], 
 \label{eq:Lifetimes_s2}
\\
t_{\tilde{Q}_1}^{(2)} & = & \frac{1}{1+x^2}\left[\frac{2x^2}{(1-x)^2}\psi(x) - \begin{cases}
       2x^2 - 2x\ln (x), & x\leq1,\\
       2 + 2x\ln (x), & x>1,
       \end{cases}\right], \nonumber\\
 t_{\tilde{Q}_2}^{(2)} & = & \frac{1}{1+x^2}\left[-\frac{x^2}{(1-x)^2}\psi(x) + \begin{cases}
       x^2 - x\ln (x), & x\leq1,\\
       1 + x\ln (x), & x>1,
       \end{cases}\right], \nonumber\\
 t_{\tilde{Q}_P}^{(2)} & = & \frac{1}{1+x^2}\left[\frac{x^2}{(1-x)^2}\psi(x) - \begin{cases}
       x^2 - x\ln (x), & x\leq1,\\
       1 + x\ln (x), & x>1,
       \end{cases}\right]. 
 \label{eq:Lifetimes_t2}
\end{eqnarray}
It can be clearly seen from Eq.(\ref{eq:Lifetimes_t2}) that the expressions for
$t^{(2)}_{\mathcal{O}}$ logarithmically diverge at the point $x=1$. For this reason, 
the weight function method is not applicable here since it requires the discontinuity
$t_\mathcal{O}^{(2)}$ to be directly evaluated at the point
$\omega_1=\omega_2=\overline{\Lambda}_s$. We briefly discuss the origin of this divergence in Appendix~\ref{sec:log_divergence}.
\\
For the condensates, we find the following expressions up to contributions of dimension six:
\begin{eqnarray}
 \Delta_\text{tree}^\text{cond}\rho_{\tilde{Q}_i^q}^q(\omega_1,\omega_2)  & = & 0 + \dots\,, \nonumber\\
 \Delta_\text{tree}^\text{cond}\rho_{\tilde{T}_1^q}^q(\omega_1,\omega_2)  & = & -\frac{\Braket{\frac{\alpha_s}{\pi}G^2}}{64\pi^2} \left(1+\frac{m_s}{\omega_1}+\frac{m_s}{\omega_2}\right)\theta(\omega_1-m_s)\,\theta(\omega_2-m_s)\nonumber\\
   & \phantom{=} & + \frac{\Braket{g_s \bar{q}\sigma_{\mu\nu}G^{\mu\nu}q}}{64\pi^2} \left[\delta(\omega_1)\,\theta(\omega_2-m_s)+\delta(\omega_2)\,\theta(\omega_1-m_s)\right] + \dots\,, \nonumber\\
 \Delta_\text{tree}^\text{cond}\rho_{\tilde{T}_2^q}^q(\omega_1,\omega_2)  & = & 0 + \dots\,, \\
 \Delta_\text{tree}^\text{cond}\rho_{\tilde{Q}_P^q}^q(\omega_1,\omega_2)  & = & \frac{\Braket{\frac{\alpha_s}{\pi}G^2}}{384\pi^2} \left(1+\frac{m_s}{\omega_1}+\frac{m_s}{\omega_2}\right)\theta(\omega_1-m_s)\,\theta(\omega_2-m_s)\nonumber\\& \phantom{=} & - \frac{\Braket{g_s \bar{q}\sigma_{\mu\nu}G^{\mu\nu}q}}{384\pi^2} \left[\delta(\omega_1)\,\theta(\omega_2-m_s)+\delta(\omega_2)\,\theta(\omega_1-m_s)\right] + \dots\,,\nonumber
 \label{eq:Lifetimes_cond}
\end{eqnarray}
from which only the bag parameters $\epsilon_1$ and $B_P$ receive non-vanishing
contributions, while 
\begin{equation}
    \Delta_\text{peng}^\text{cond}\rho_{\tilde{Q}_i^{q'}}^q(\omega_1,\omega_2)   =  0 + \dots\,,
\end{equation}
as discussed above and therefore there are no condensate corrections to the $\delta$s at this order. Considering the case $m_s=0$ we find perfect agreement with the
results found in Ref.~\cite{Cheng:1998ia}\footnote{The additional factor of 4 appearing in Eq.(3.24) of Ref.~\cite{Cheng:1998ia} is accounted for by their choice of operator normalisation.}. The analysis by 
Ref.~\cite{Baek:1998vk} chooses instead an axial-vector interpolating current, $\overline{q}\gamma_{\alpha}\gamma^5h$, and therefore their results differ from our own
in addition to the inconsistency mentioned in Section \ref{subsec:condensates}. 
As pointed out in Ref.~\cite{Cheng:1998ia}, this choice means that
states of quantum number $J^P=1^+$ are also being considered by the correlation function.


\section{Results}
\label{sec:numerics} 
\begin{figure}
     \centering
     \begin{subfigure}[b]{0.4\textwidth}
         \centering
         \includegraphics[width=\textwidth]{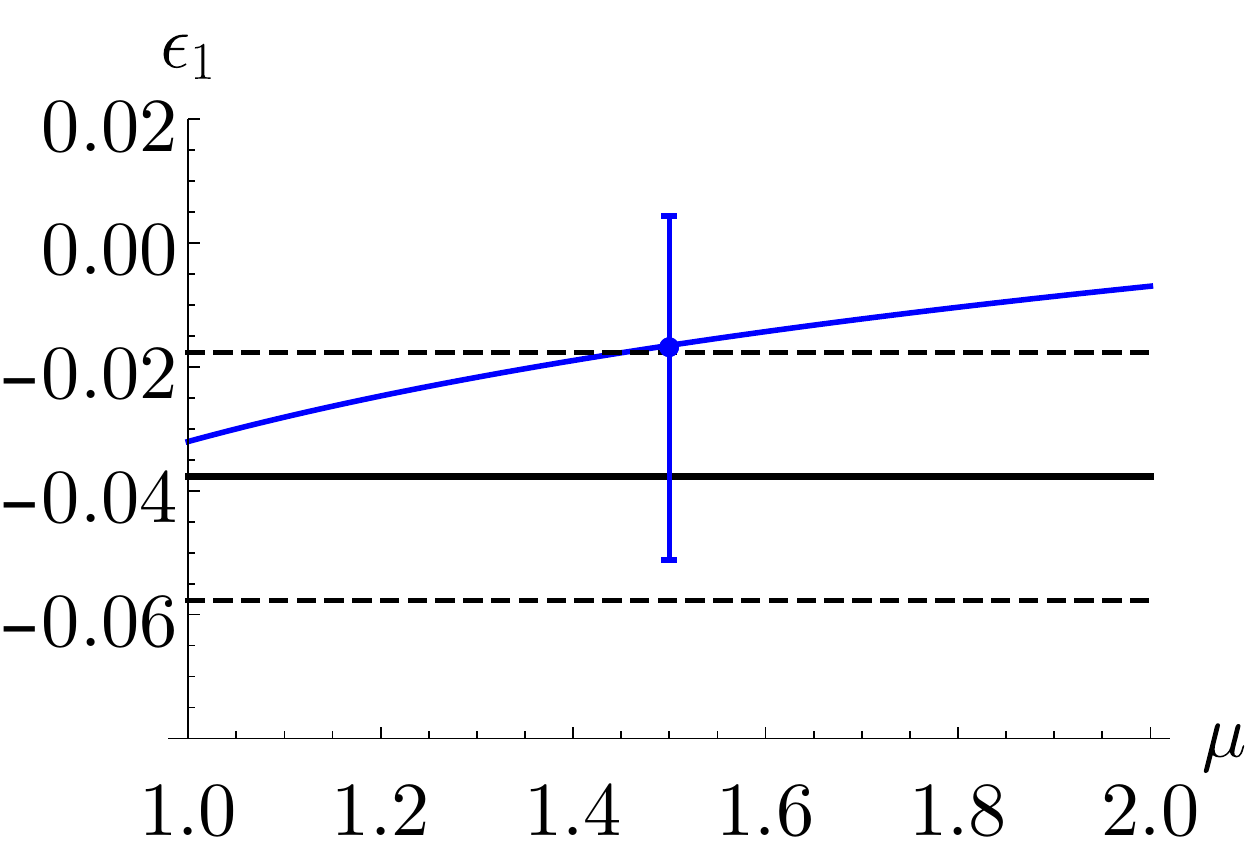}
         \caption{}
     \end{subfigure}
     \hfill
     \begin{subfigure}[b]{0.4\textwidth}
         \centering
         \includegraphics[width=\textwidth]{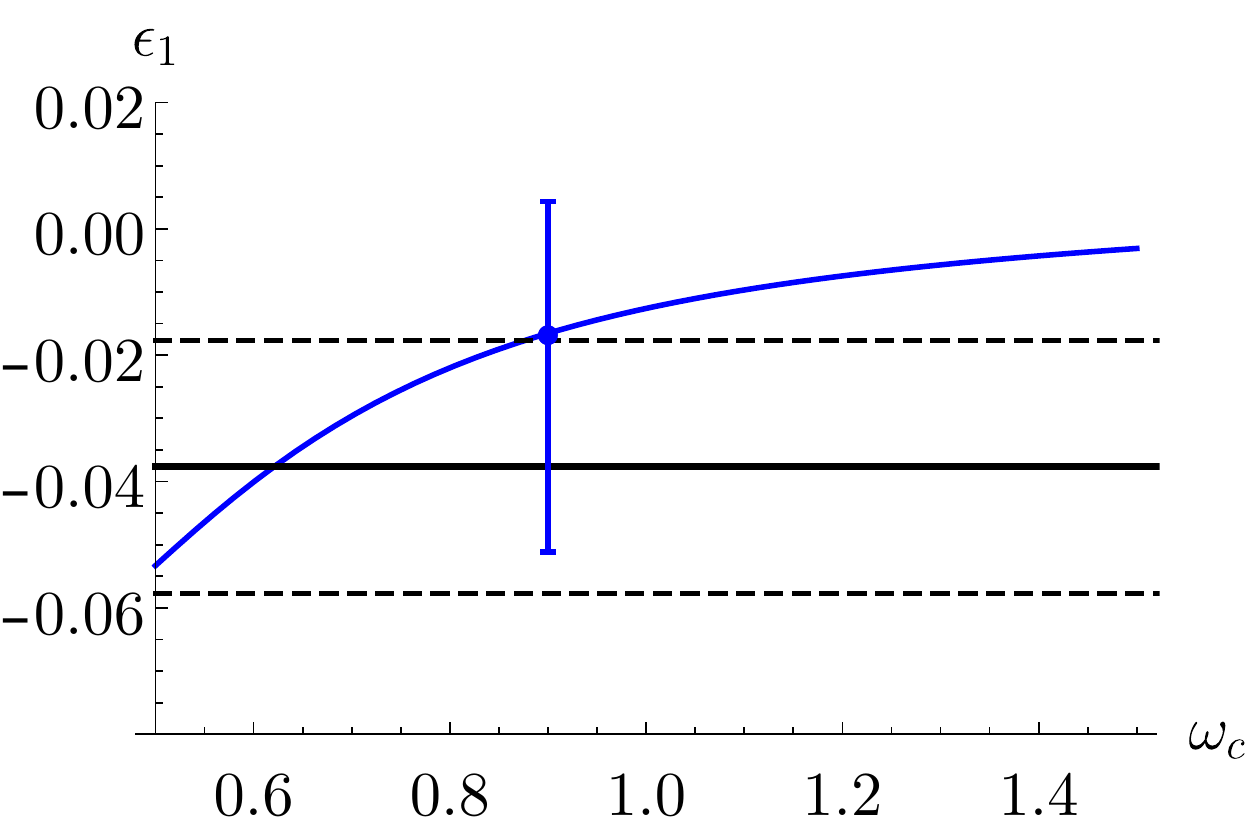}
         \caption{}
     \end{subfigure}
     \vspace{0.5mm}
     \begin{subfigure}[b]{0.4\textwidth}
         \centering
         \includegraphics[width=\textwidth]{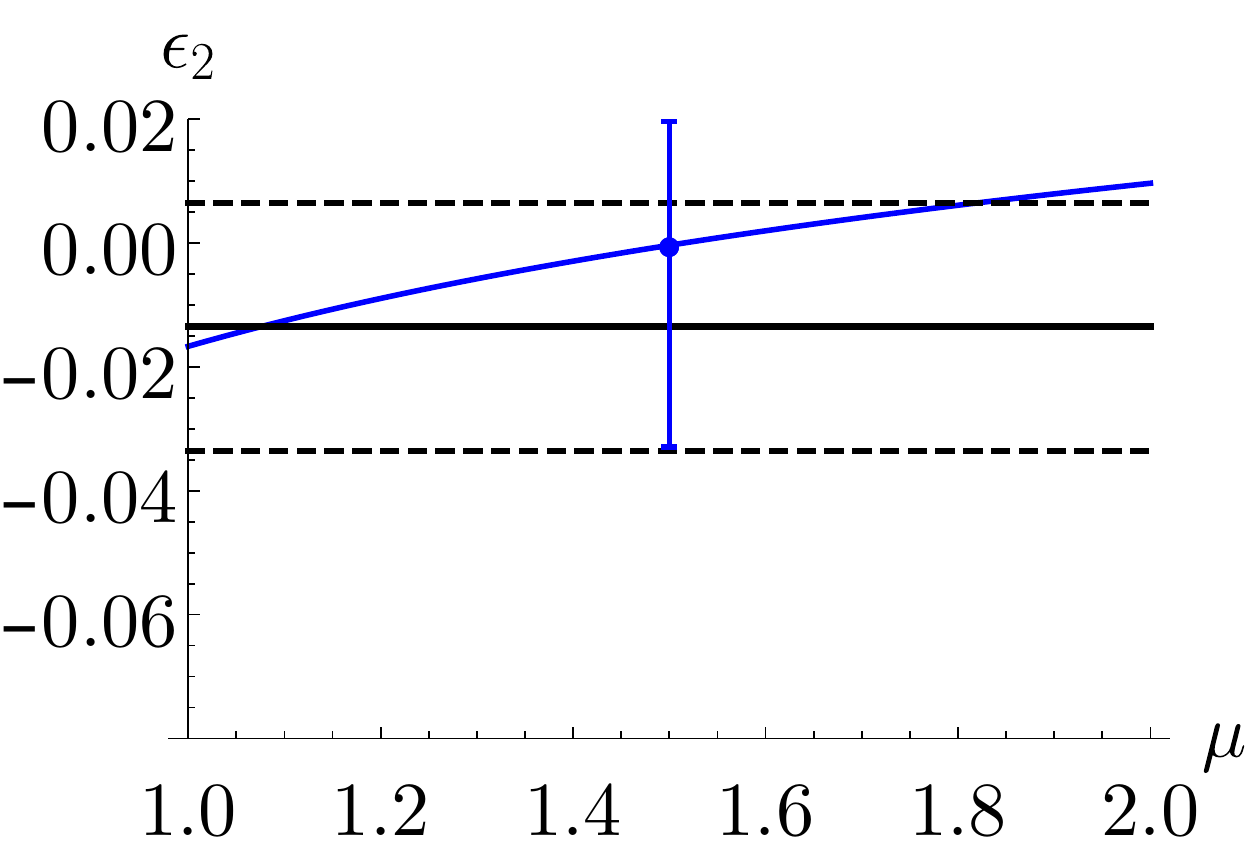}
         \caption{}
     \end{subfigure}
     \hfill
     \begin{subfigure}[b]{0.4\textwidth}
         \centering
         \includegraphics[width=\textwidth]{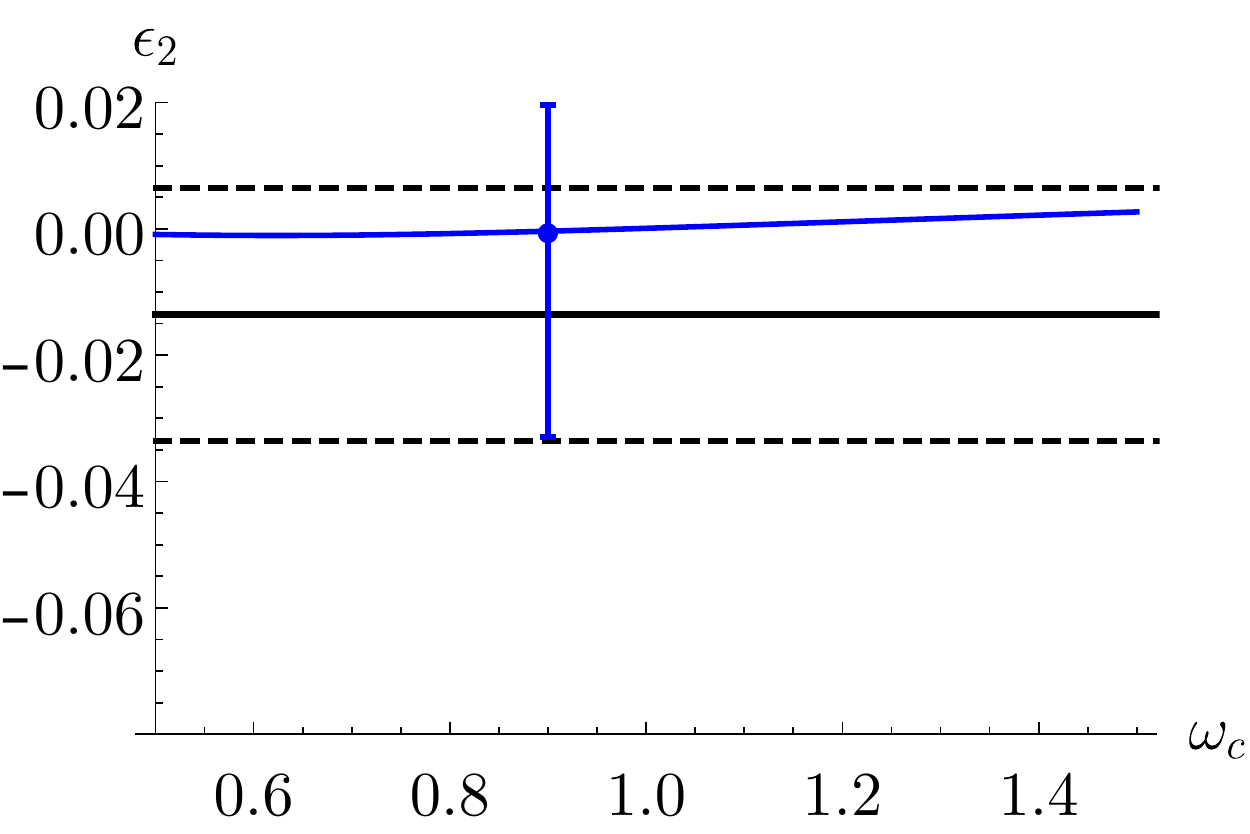}
         \caption{}
     \end{subfigure}
     \begin{subfigure}[b]{0.4\textwidth}
         \centering
         \includegraphics[width=\textwidth]{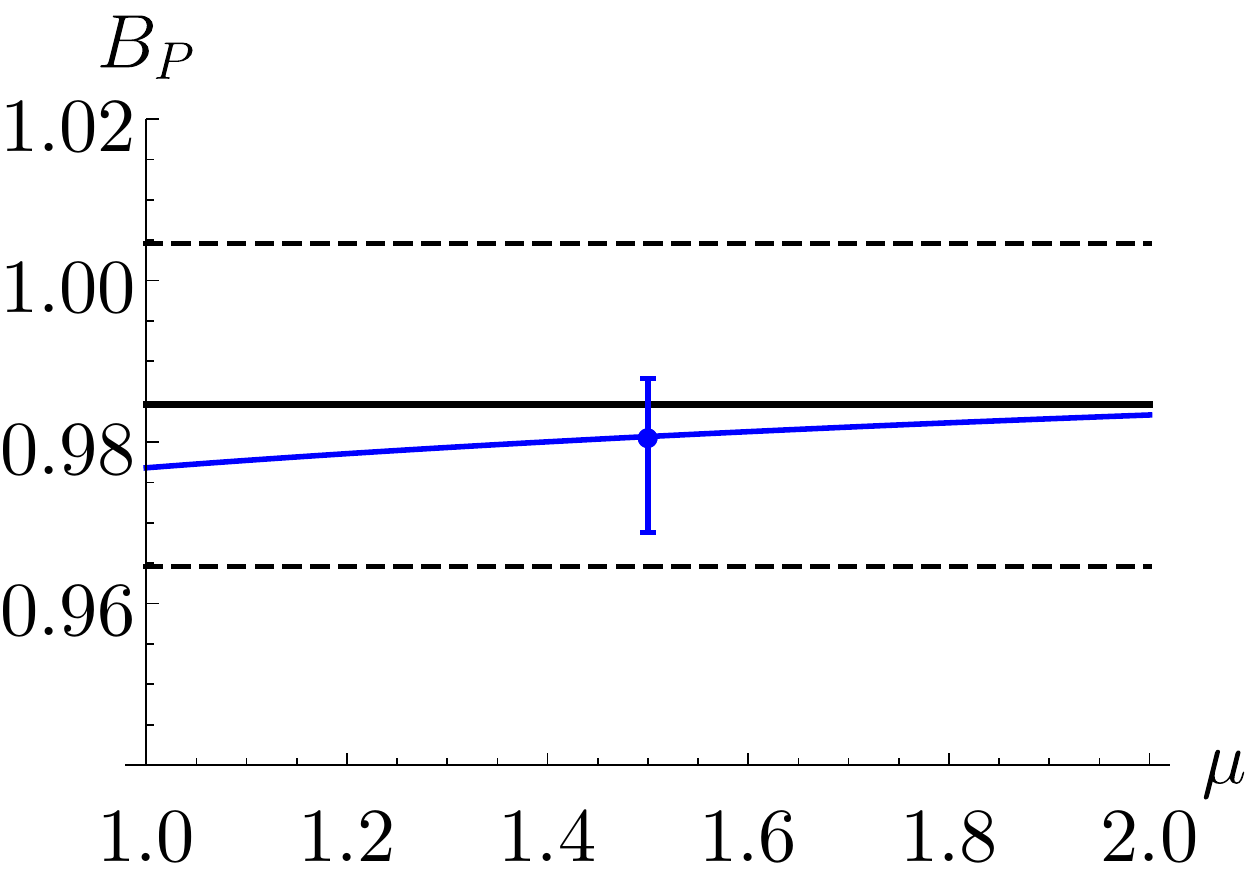}
         \caption{}
     \end{subfigure}
     \hfill
     \begin{subfigure}[b]{0.4\textwidth}
         \centering
         \includegraphics[width=\textwidth]{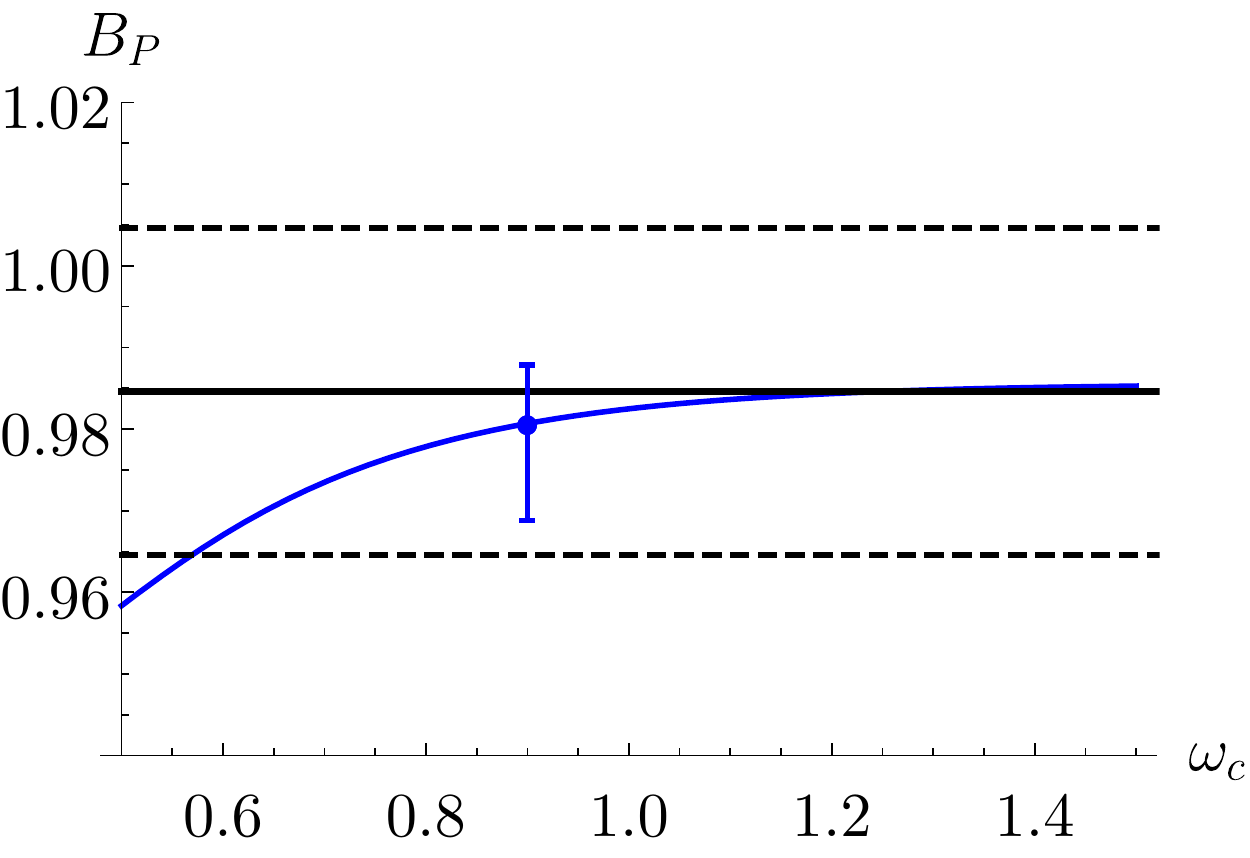}
         \caption{}
     \end{subfigure}
     \caption{Comparison of the weight function method (shown in black) to the traditional sum rule approach (shown in blue) for the case of a $B_d$ meson. The plots illustrate how the traditional result varies with respect to $\mu$ and $\omega_c$ on the left and right respectively. The dashed lines indicate the range of uncertainty in the weight function result, being set to $\pm0.02$. The blue vertical line indicates our final quoted error for the traditional sum rule method.}
     \label{fig:WFM-TSR}
\end{figure}

Two methods of carrying out the sum rule are available to us: the
weight-function-method described in Ref.~\cite{Kirk:2017juj} and the
traditional sum rule approach in which we explicitly evaluate
Eq.(\ref{eq:SumRuleVal}) and Eq.(\ref{eq:SumRuleNonVal}). Since having a
non-zero strange quark mass in the eye contraction terms and the use of the
weight function method are incompatible with one another, we choose to use a
traditional approach for the main numerical results presented in this section. However, where applicable a direct comparison of both methods was also carried
out in which we find a reassuring level of consistency, see Fig.~\ref{fig:WFM-TSR}.

In our analysis the continuum cut-off $\omega_c$, and the Borel parameter $t$ are fixed for the cases of the $B_d$ (because of isospin in our analysis $B_u = B_d$) and $B_s$ mesons separately through a sum rule analysis of their respective decay constants and mass differences. From this analysis we find,
\begin{align}
 B_d: \;\; w_c = 0.90\,\text{GeV} ,\;\; t = 1\,\text{GeV}, \\
 B_s: \;\; w_c = 0.95\,\text{GeV} ,\;\; t = 1\,\text{GeV}. 
\end{align}
\\
We evaluate the sum rules for the HQET bag parameters at the scale $\mu=1.5$\,GeV.
For the strange quark mass, we use the $\overline{\text{MS}}$ scheme value at the
scale $\mu=1.5$\,GeV after running \cite{Herren:2017osy} from $\overline{m_s}$(2GeV)$=95^{+9}_{-3}$\,MeV. 
As in the analysis of Ref.~\cite{King:2019lal}, we expand the range of uncertainty
to $95\pm30$\,MeV in order to account for the missing terms after our truncation of
the $m_s$ expansion and scheme dependencies. After inspecting the range of stability
in the HQET sum rules of $F_{d/s}$ and $\overline{\Lambda}$, we chose to vary $t$ by
$\pm0.4$ and to vary $\omega_c$ by $\pm0.2$ in our error analysis. 
The uncertainty associated with the sum rule scale is estimated by varying $\mu$
between 1-2\,GeV, running back to the central value of 1.5\,GeV and then
scaling\footnote{We believe this treatment is justified given the usual procedure of
varying between $[\mu/2,2\mu]$ is not practical at such low scales, and so re-scale
the uncertainty in order to compensate for this limitation.} the resulting uncertainty
by a factor of 2. 
A list of the other parameters used in this work is presented in Table\,\ref{tab:Inputs}
and
includes the values used for the condensates which are quoted at the scale 2\,GeV. 
\\
We use the relation, $\Braket{g_s \bar{q}\sigma_{\mu\nu}G^{\mu\nu}q}=m_0^2\Braket{\overline{q}q}$ 
at the scale 2\,GeV 
with $m_0^2=0.8$\,GeV$^2$ \cite{Belyaev:1982sa} in order to determine the value of the
mixed quark-gluon condensate. The renormalisation group
equations describing the running of the condensates down to the sum rule scale can be found in Appendix\,\ref{sec:FandLambda_SR}. In our analysis, a more conservative estimate for their individual uncertainties of $\pm30\%$ was chosen over the values quoted in Table \ref{tab:Inputs} in order to account for the accuracy in $m_0^2$.
\begin{table}
    \centering
    \begin{tabular}{l c r l c}
    \toprule
        $\overline{m_s}$(2GeV) &\phantom{-}& $95^{+9}_{-3}$  & MeV & \cite{ParticleDataGroup:2018ovx} \\
    \midrule[0.2pt] 
        $\braket{\frac{\alpha_s}{\pi}G^2}$ &\phantom{-}& $0.012\pm0.006$ & GeV$^4$ & \cite{Ball:2006wn} \\ 
    \midrule[0.2pt]
        $\braket{\overline{d}d}$(2GeV) &\phantom{-}& $(-0.283\pm0.002)^3$ & GeV$^3$ & \cite{McNeile:2012xh} \\ 
    \midrule[0.2pt]
        $\braket{\overline{s}s}$(2GeV) &\phantom{-}& $(-0.296\pm0.002)^3$ & GeV$^3$ & \cite{Davies:2018hmw} \\ 
    \midrule[0.2pt]
        $\overline{m_b}(\overline{m_b})$ &\phantom{-}& $4.203^{+0.016}_{-0.034}$ & GeV & \cite{Beneke:2014pta,Beneke:2016oox} \\ 
    \midrule[0.2pt]
        $M_Z$ &\phantom{-}& $91.1876$ & GeV & \cite{Zyla:2020zbs} \\ 
    \midrule[0.2pt]
        $\alpha_s(M_Z)$ &\phantom{-}& $0.1181\pm 0.0011$ &\phantom{-}& \cite{ParticleDataGroup:2018ovx} \\ 
    \bottomrule
    \end{tabular}
    \caption{Values of input parameter used in our numerical analysis.}
    \label{tab:Inputs}
\end{table}

Our numerical results for the bag parameters $B_i$, $\epsilon_i$ and $B_P$ for
the $B^d$ and $B^s$ systems can be found in Table\,\ref{tab:B_d_results} and
Table\,\ref{tab:B_s_results} respectively, where the total estimated
uncertainty is denoted by $\alpha$. The contribution to the uncertainty
associated with variations of the sum rule scale is denoted by $\alpha_\mu$,
whereas $\alpha_P$ represents the combined parametric uncertainty of $m_s$, the
Borel parameter, the sum rule cut-off, and the condensates. We stress again
that these parameters are taken in the strict HQET limit $m_b\rightarrow\infty$
and therefore we do not quote an uncertainty associated with $1/m_b$
corrections.\\ 
Evidently, the dominant source of uncertainty arises from scale variations. The
parametric uncertainty seems negligible in comparison, with the exception of
$\epsilon_1$ and $B_P$. Unlike the other bag parameters, these receive
non-vanishing condensate contributions (see Eq.(\ref{eq:Lifetimes_cond})) and as
a consequence are found to have a greater dependence on the cut-off $\omega_c$
and are sensitive to the numerical input of the condensates themselves. It should
be noted that in our analysis we found that dependence on the Borel parameter was
weak\footnote{As was also found to be the case in Ref.~\cite{Cheng:1998ia}.}.\\

\begin{table}
    \centering
    \begin{tabular}{C C C C C C C C}
        \toprule
        B_i^d & \text{TSR} & \alpha & \phantom{-}\mathcal{O}(m_d^0) & \phantom{-}\mathcal{O}(m_d^1) & \phantom{-}\mathcal{O}(m_d^2) & \alpha_\mu & \alpha_P \\
        \toprule
        B_1^d & \phantom{-}1.0026 & ^{+0.0198}_{-0.0106} & \phantom{-}1.0026 & - & - & ^{+0.0197}_{-0.0105} & ^{+0.0005}_{-0.0007} \\
        \midrule[0.2pt]
        B_2^d & \phantom{-}0.9982 & ^{+0.0052}_{-0.0066} & \phantom{-}0.9982 & - & - & ^{+0.0051}_{-0.0066} & ^{+0.0005}_{-0.0004} \\
        \midrule[0.2pt]
        \epsilon_1^d & -0.0165 & ^{+0.0209}_{-0.0346} & -0.0165 & - & - & ^{+0.0191}_{-0.0310} & ^{+0.0084}_{-0.0153} \\
        \midrule[0.2pt]
        \epsilon_2^d & -0.0004 & ^{+0.0200}_{-0.0326} & -0.0004 & - & - & ^{+0.0200}_{-0.0326} & ^{+0.0010}_{-0.0006} \\
        \midrule[0.2pt]
        B_P^d & \phantom{-}0.9807 & ^{+0.0072}_{-0.0119} & \phantom{-}0.9807 & - & - & ^{+0.0053}_{-0.0077} & ^{+0.0049}_{-0.0091} \\
        \bottomrule
    \end{tabular}
    \caption{Bag parameter results for the $B_d$ system using the traditional sum rule `TSR'.}
    \label{tab:B_d_results}
\end{table}
\begin{table}
    \centering
    \begin{tabular}{C C C C C C C C}
        \toprule
        B_i^s & \text{TSR} & \alpha & \phantom{-}\mathcal{O}(m_s^0) & \phantom{-}\mathcal{O}(m_s^1) & \phantom{-}\mathcal{O}(m_s^2) & \alpha_\mu & \alpha_P \\
        \toprule
        B_1^s & \phantom{-}1.0022 & ^{+0.0185}_{-0.0099} & \phantom{-}1.0019 & \phantom{-}0.0006 & - 0.0003 & ^{+0.0185}_{-0.0099} & ^{+0.0004}_{-0.0005} \\
        \midrule[0.2pt]
        B_2^s & \phantom{-}0.9983 & ^{+0.0052}_{-0.0067} & \phantom{-}0.9986 & - 0.0004 & \phantom{-}0.0001 & ^{+0.0052}_{-0.0067} & ^{+0.0004}_{-0.0003} \\
        \midrule[0.2pt]
        \epsilon_1^s & -0.0104 & ^{+0.0202}_{-0.0330} & -0.0097 & - 0.0008 & \phantom{-}0.0002 & ^{+0.0195}_{-0.0319} & ^{+0.0051}_{-0.0084} \\
        \midrule[0.2pt]
        \epsilon_2^s & \phantom{-}0.0001  & ^{+0.0199}_{-0.0324} & -0.0001 & \phantom{-}0.0002 & \phantom{-} 0.0001  & ^{+0.0199}_{-0.0324} & ^{+0.0010}_{-0.0008} \\
        \midrule[0.2pt]
        B_P^s & \phantom{-}0.9895 & ^{+0.0053}_{-0.0077} & \phantom{-}0.9873 & \phantom{-}0.0016 & \phantom{-}0.0006 & ^{+0.0043}_{-0.0059} & ^{+0.0031}_{-0.0050} \\
        \bottomrule
    \end{tabular}
    \caption{Bag parameter results for the $B_s$ system using the traditional sum rule `TSR'.}
    \label{tab:B_s_results}
\end{table}
We find very good convergence properties in the $m_s$ expansion suggesting that
we can be confident in the validity of the `expansion by regions' method and in
a sufficient accuracy when working up to order $m_s^2$. 
Numerical differences between the $\mathcal{O}(m_s^0)$ term of
the $B^s$ bag parameters and those of $B^d$ come from 3 sources: different
input for the condensates, the lower cut of the sum rule integral (see
Eq.(\ref{eq:SumRuleVal}) and Eq.(\ref{eq:SumRuleNonVal})), and a different value of the decay constant in the denominator since we do not expand the ratio in $m_s$.
\\
At NLO in $\alpha_s$, the only contribution to the bag parameters of the colour
singlet operators comes from eye-contraction diagrams and therefore the deviation
from their VSA value is suppressed in comparison to the bag parameters for the
colour octet and penguin operators. 
\\
Our numerical findings for the non-valence bag parameters are presented in
Tables\,\ref{tab:delta_dd}-\ref{tab:delta_sd}. Again no
significant shift away from the VSA values was found.
Additionally, flavour breaking effects in the form of $m_s$
corrections are small. The first
non-vanishing corrections from the strange quark mass in
the operator of an eye contraction diagram appear at
$\mathcal{O}(m_s^2)$. This corresponds with the results for
$\delta_i^{sd}$ shown in Table\,\ref{tab:delta_sd}.
\begin{table}
    \centering
    \begin{tabular}{C C C C C C C C}
        \toprule
        \delta_i^{ud} & \text{TSR} & \alpha & \phantom{-}\mathcal{O}(m_d^0) & \phantom{-}\mathcal{O}(m_d^1) & \phantom{-}\mathcal{O}(m_d^2) & \alpha_\mu & \alpha_P \\
        \toprule
        \delta_1^{ud} & \phantom{-}0.0026 & ^{+0.0142}_{-0.0092} & \phantom{-}0.0026 & - & - & ^{+0.0142}_{-0.0092} & ^{+0.0005}_{-0.0007} \\
        \midrule[0.2pt]
        \delta_2^{ud} & -0.0018 & ^{+0.0047}_{-0.0072} & -0.0018 & - & - & ^{+0.0046}_{-0.0071} & ^{+0.0005}_{-0.0004} \\
        \midrule[0.2pt]
        \delta_3^{ud} & -0.0004 & ^{+0.0015}_{-0.0024} & -0.0004 & - & - & ^{+0.0015}_{-0.0024} & ^{+0.0001}_{-0.0001} \\
        \midrule[0.2pt]
        \delta_4^{ud} & \phantom{-}0.0003 & ^{+0.0012}_{-0.0008} & \phantom{-}0.0003 & - & - & ^{+0.0012}_{-0.0008} & ^{+0.0001}_{-0.0001} \\
        \midrule[0.2pt]
        \delta_P^{ud} & -0.0083 & ^{+0.0209}_{-0.0322} & -0.0083 & - & - & ^{+0.0208}_{-0.0322} & ^{+0.0025}_{-0.0017} \\
        \bottomrule
    \end{tabular}
    \caption{Non-valence bag parameters for the case $q=q'=u,d$ 
    (note $\delta^{ud}=\delta^{du}$ ) using the traditional sum rule `TSR'.}
    \label{tab:delta_dd}
\end{table}
\begin{table}
    \centering
    \begin{tabular}{C C C C C C C C}
        \toprule
        \delta_i^{ds} & \text{TSR} & \alpha & \phantom{-}\mathcal{O}(m_s^0) & \phantom{-}\mathcal{O}(m_s^1) & \phantom{-}\mathcal{O}(m_s^2) & \alpha_\mu & \alpha_P \\
        \toprule
        \delta_1^{ds} & \phantom{-}0.0025 & ^{+0.0144}_{-0.0093} & \phantom{-}0.0019 & \phantom{-}0.0006 & - 0.0000 & ^{+0.0144}_{-0.0093} & ^{+0.0004}_{-0.0005} \\
        \midrule[0.2pt]
        \delta_2^{ds} & -0.0018 & ^{+0.0047}_{-0.0072} & -0.0014 & - 0.0004 & \phantom{-}0.0000 & ^{+0.0047}_{-0.0072} & ^{+0.0004}_{-0.0003} \\
        \midrule[0.2pt]
        \delta_3^{ds} & -0.0004 & ^{+0.0015}_{-0.0024} & -0.0003 & - 0.0001 & \phantom{-}0.0000 & ^{+0.0015}_{-0.0024} & ^{+0.0001}_{-0.0001} \\
        \midrule[0.2pt]
        \delta_4^{ds} & \phantom{-}0.0003 & ^{+0.0012}_{-0.0008} & \phantom{-}0.0002 & \phantom{-}0.0001 & - 0.0000 & ^{+0.0012}_{-0.0008} & ^{+0.0001}_{-0.0001} \\
        \midrule[0.2pt]
        \delta_P^{ds} & -0.0041 & ^{+0.0217}_{-0.0338} & -0.0062 & \phantom{-}0.0020 & \phantom{-}0.0001 & ^{+0.0217}_{-0.0338} & ^{+0.0018}_{-0.0015} \\
        \bottomrule
    \end{tabular}
    \caption{Non-valence bag parameters with a strange spectator quark using the traditional sum rule `TSR'.}
    \label{tab:delta_ds}
\end{table}

\begin{table}
    \centering
    \begin{tabular}{C C C C C C C C}
        \toprule
        \delta_i^{sd} & \text{TSR} & \alpha & \phantom{-}\mathcal{O}(m_s^0) & \phantom{-}\mathcal{O}(m_s^1) & \phantom{-}\mathcal{O}(m_s^2) & \alpha_\mu & \alpha_P \\
        \toprule
        \delta_1^{sd} & \phantom{-}0.0023 & ^{+0.0140}_{-0.0091} & \phantom{-}0.0026 & - & - 0.0004 & ^{+0.0140}_{-0.0090} & ^{+0.0005}_{-0.0007} \\
        \midrule[0.2pt]
        \delta_2^{sd} & -0.0017 & ^{+0.0046}_{-0.0070} & -0.0018 & - & \phantom{-}0.0002 & ^{+0.0046}_{-0.0070} & ^{+0.0006}_{-0.0004} \\
        \midrule[0.2pt]
        \delta_3^{sd} & -0.0004 & ^{+0.0015}_{-0.0023} & -0.0004 & - & \phantom{-}0.0001 & ^{+0.0015}_{-0.0023} & ^{+0.0001}_{-0.0001} \\
        \midrule[0.2pt]
        \delta_4^{sd} & \phantom{-}0.0003 & ^{+0.0012}_{-0.0008} & \phantom{-}0.0003 & - & - 0.0000 & ^{+0.0012}_{-0.0008} & ^{+0.0001}_{-0.0001} \\
        \midrule[0.2pt]
        \delta_P^{sd} & -0.0074 & ^{+0.0207}_{-0.0316} & -0.0083 & - & \phantom{-}0.0008 & ^{+0.0205}_{-0.0315} & ^{+0.0025}_{-0.0017} \\
        \bottomrule
    \end{tabular}
    \caption{Non-valence bag parameters considering a strange light quark in the operator using the traditional sum rule `TSR'.}
    \label{tab:delta_sd}
\end{table}

The plots in Fig.\,\ref{fig:WFM-TSR}
show the dependence of the colour octet and penguin bag parameters on the
sum rule scale and the continuum cutoff for the
$B_d$ meson as calculated using the traditional
sum rule method. Also indicated on the plots is an alternative result for which the perturbative tree contribution has been evaluated using the weight function
analysis\footnote{The corresponding plots for the colour singlet bag parameters have been omitted since they do not receive contributions from tree contraction terms.}. Comparing the two methods we observe
that the predictions lie within the range of 
uncertainties of each other and therefore
demonstrate a sound level of consistency which provides us with further confidence in the validity of the results presented in this paper.

Finally we can  compare our results with other 
sum rule analyses of the bag parameters that are 
available in the literature. The treatment in Ref.~\cite{Cheng:1998ia} shows several key
differences compared to ours: in that study, the
necessary tools to calculate the dominant
perturbative 3-loop non-factorisable contributions
shown in Fig.\,\ref{fig:3loops} were not yet
available. However, additional non-factorisable
effects do arise from their procedure for
extracting the continuum cut-off, which in their
case is not treated as common between the 3-point
and 2-point correlators. The main result of that
paper is quoted at the scale $m_b$, for which
there is significant mixing between the bag
parameters after running from the hadronic scale. It should also be noted that their results differ
from our own by a factor of $F^2(m_b)/F^2(\mu)$
due to different conventions in our definition of the matrix elements, (see Eq.(\ref{eq:Matrix_Elements})).

The latest preliminary estimates  of the lifetime bag parameters with lattice
QCD were obtained 20 years ago in Ref.~\cite{Becirevic:2001fy} and so an updated
analysis would be greatly appreciated. Comparing those values to our own we find
a similar degree of precision for the $\epsilon_i$ parameters, while our
predictions for the $B_i$ have a much smaller range of uncertainty and we
disagree with the low value quoted for $B_2$.


\section{Conclusion}
\label{sec:conclusions} 
In this work, we have presented an updated sum rule analysis of the $\Delta B=0$ bag parameters in the HQET limit which includes SU(3) flavour breaking effects for the first time, relying on the expansion by regions approach we introduced in our earlier work \cite{King:2019lal} in the context of $B$-meson mixing. The presence of the eye-contraction diagrams and the mixing between operators of different dimensions in full QCD however poses an additional challenge. For this reason, we work exclusively in HQET where no such mixing occurs. Therefore, the results presented here are also
applicable to the $\Delta D=0$ matrix elements (see Ref.~\cite{King:2021xqp} 
for a recent update of $D$ meson lifetimes). In addition, taking this limit
leads us to find relatively small uncertainties for the bag parameters
themselves since all $1/m_Q$ corrections reside in $\tilde{\Gamma}_7 $ of
Eq.(\ref{eq:HQE}).  

The eye contractions are first addressed in this work and also lead to a number of new effects. First of all, their renormalization requires the inclusion of the penguin operator $Q_P$ in our operator basis. Furthermore, since the light-quarks $q'$ in the operators are not contracted with the light valence quarks $q$ in the mesons, they generate non-valence matrix elements $\delta_i^{q'q}$ for $q\neq q'$. We find that the weight-function method we employed in \cite{King:2019lal} cannot be used with the non-valence matrix elements due to logarithmic divergences whose origin is discussed in Appendix~\ref{sec:log_divergence}. Thus, we adopted the traditional sum rule approach where the Borel parameter and the continuum cutoff are varied in our analysis. We note however that we obtain good consistency between the two methods when they are applied to the tree contractions as shown in Figure~\ref{fig:WFM-TSR}. 

Numerically, we find that deviations from the VSA at the hadronic scale are generally small. The $\mathcal{O}(1)$ uncertainties in the sum rule for the deviations are therefore quite small in absolute terms and sufficient for a phenomenological analysis of the $\tau(B_s)/\tau(B_d)$ lifetime ratio, which also requires taking into account the contribution of the Darwin operator \cite{Lenz:2020oce,Mannel:2020fts,MorenoTorres:2020xir} and is thus beyond the scope of this work.

\section*{Acknowledgments}
The work of D.K.  was supported by the STFC grant of the
IPPP. A.L. would like to thank Maria Laura Piscopo for proof-reading of the manuscript. 

\appendix
\section{RGE}
\label{sec:RGE}
To determine the counterterm contribution to the three-point correlator \eqref{eq:DefK} we require
the one-loop renormalization of the operators \eqref{eq:Lifetimes_HQET_operators}. 
We obtain the structure

\begin{equation}
\gamma_{\tilde{O}^{q'}\tilde{O}^q} = \delta_{qq'}\gamma_{\tilde{O}\tilde{O}} + \gamma_{\tilde{O}'\tilde{O}}
\end{equation}

with

\begin{equation}
\tilde{\gamma}_{\tilde{O}\tilde{O}}^{(0)} = \left(
\begin{array}{ccccc}

  \frac{3}{N_c}-3 N_c & 0 & 6 & 0 & 0\\ \vspace*{0.1cm}

0 & \frac{3}{N_c}-3 N_c & 0 & 6 & 0\\ \vspace*{0.1cm}

\frac{3}{2}-\frac{3}{2 N_c^2} & 0 & -\frac{3}{N_c} & 0 & 0\\ \vspace*{0.1cm}

0 & \frac{3}{2}-\frac{3}{2 N_c^2} & 0 & -\frac{3}{N_c} & 0\\ \vspace*{0.1cm}

0 & 0 & 0 & 0 & -3N_c
\end{array}
\right),
\label{eq:lifetimes_ADM_OtilOtil}
\end{equation}

and

\begin{equation}
\tilde{\gamma}_{\tilde{O}'\tilde{O}}^{(0)} = \left(
\begin{array}{ccccc}

0 & 0 & 0 & 0 & \frac{8}{3}\\ \vspace*{0.1cm}

0 & 0 & 0 & 0 & -\frac{4}{3}\\ \vspace*{0.1cm}

0 & 0 & 0 & 0 & -\frac{4}{3N_c}\\ \vspace*{0.1cm}

0 & 0 & 0 & 0 & \frac{2}{3N_c}\\ \vspace*{0.1cm}

0 & 0 & 0 & 0 & \frac{4}{3}

\end{array}
\right),
\label{eq:lifetimes_ADM_OtilOtilprime}
\end{equation}

The renormalized correlator then takes the form

\begin{equation}
K_{\tilde{Q}_i^{q'}}^{q,(1)} = K_{\tilde{Q}_i^{q'}}^{q,(1),\text{bare}} + \frac{1}{2\epsilon}\left[  \left(2\tilde{\gamma}_{\tilde{j}}^{(0)}\delta_{ij}+\tilde{\gamma}_{\tilde{Q}_i\tilde{Q}_j}^{(0)}\right)K_{\tilde{Q}_j^{q'}}^{q,(0)}+\tilde{\gamma}_{\tilde{Q}_i\tilde{E}_j}^{(0)}K_{\tilde{E}_j^{q'}}^{q,(0)}\right]+ \frac{1}{2\epsilon}\tilde{\gamma}_{\tilde{Q}_i'\tilde{Q}_{P}}^{(0)}K_{\tilde{Q}_{P}^{q}}^{q,(0)},
\end{equation}

where the second term is the counterterm for the tree-level contractions and the third term is the counterterm for the eye contractions.

Now, we consider the RGE for the Bag parameters. We have

\begin{equation}
\frac{d\tilde{\mathbf{O}}^{q'}}{d \ln\mu} = -\sum_q \tilde{\gamma}_{\tilde{O}^{q'}\tilde{O}^q} \tilde{\mathbf{O}}^{q}\,, \hspace{1cm} \frac{dF_q(\mu)}{d \ln\mu} = -\tilde{\gamma}_{\tilde{j}} F_q(\mu)\,,
\end{equation}

and thus obtain the following RGE for the Bag parameters in the case with two light-quark flavors $q$ and $s$:

\begin{equation}
   \frac{d}{d\ln\mu}\left(\begin{array}{c}\tilde{\mathcal{B}}_i^q\\\tilde{\delta}_i^{qs}\end{array}\right)
 = -\frac{\tilde{A}_j}{\tilde{A}_i}
   \left(\begin{array}{cc}
   \tilde{\gamma}_{\tilde{O}_i\tilde{O}_j} + \tilde{\gamma}_{\tilde{O}_i'\tilde{O}_j} - 2\tilde{\gamma}_{\tilde{j}}\delta_{ij} & \tilde{\gamma}_{\tilde{O}_i'\tilde{O}_j}\\

   \tilde{\gamma}_{\tilde{O}_i'\tilde{O}_j} & \tilde{\gamma}_{\tilde{O}_i'\tilde{O}_j} - 2\tilde{\gamma}_{\tilde{j}}\delta_{ij}
   \end{array}\right)
   \left(\begin{array}{c}\tilde{\mathcal{B}}_j^q\\\tilde{\delta}_j^{qs}\end{array}\right)\,,
\end{equation}
which can be easily generalised to more than two quark flavours.

\section{Condensate Calculation}
\label{sec:CondensateExample}
Here we lay out as an example our steps to derive the double discontinuity of the non-factorisable gluon-gluon condensate term. Specifically, this example concerns the case of the 3-point function with a penguin operator insertion but, other than alterations to the Dirac structure stemming from the choice of operator, the process is identical for the rest of the operator basis. For the sake of brevity, we take the case of a massless light quark. Following from the procedure described in Ref.~\cite{Novikov:1983gd} we work in the Fock-Schwinger gauge \cite{Fock:1937dy,Schwinger:1951nm},
\begin{equation}
    (x-x_0)A_{\mu}^a=0
\end{equation}
where $x_0$ can be set to zero without loss of generality as its dependence drops out of any gauge-invariant quantity. After Wick contracting the fields, the correlator in Eq.(\ref{eq:DefK}) takes the form,
\begin{equation}
  K_{\tilde{Q}_P}(\omega_1,\omega_2)=  -\int [dk]\frac{\text{Tr}\left[\gamma^5(1+\slashed{v})\gamma_{\mu}(1+\slashed{v})\gamma^5 S_{ij}(-k_2)\gamma^{\mu}S_{kl}(-k_1)\right]}{(-2(k_1\cdot v+\omega_1))(-2(k_2\cdot v+\omega_1))}\,T^a_{jk}T^a_{li}
\label{eq:condensate_corellator}
\end{equation}
where we define our integral measure as $[dk]\equiv d^dk_1d^dk_2/(2\pi)^{2d}$ and it is explicit that we choose to work in momentum space. In Eq.(\ref{eq:condensate_corellator}) we have ignored the contribution from the eye-contraction term since condensate corrections to such diagrams are vanishing at the order considered in this paper (see discussion in Section~\ref{subsec:condensates}). Furthermore, in Eq.(\ref{eq:condensate_corellator}) and in what follows, we drop the notation indicating the flavour of the light quarks appearing in the operator and in the pseudoscalar currents since for this example we take $q=q'$ and without mass corrections the result for $u/d$ is identical. The appearance of the gluon-gluon condensate arises from the next to leading order terms in the expansion of the the light quark propagators,
\begin{equation}
\begin{split}
    S_{ij}(-k) &= S^{(0)}(-k)\delta_{ij}+\frac{i g}{2}\int d^4 p\,  S^{(0)}(-k) G^b_{\rho \alpha}(0)T^b_{ij}\gamma^{\alpha}\frac{\partial}{\partial p_{\rho}}\delta^{(4)}(p)S^{(0)}(-k-p)+\mathcal{O}(g^2)\\
    &\simeq S^{(0)}(-k)+\frac{i g}{2}S^{(0)}(-k) G^b_{\rho \alpha}(0)T^b_{ij}\gamma^{\alpha} \left. \frac{\partial S^{(0)}(-k-p)}{\partial p_{\rho}}\right |_{p=0}
\end{split}
\label{eq:light_quark_propagator}
\end{equation}
Inserting Eq.(\ref{eq:light_quark_propagator}) into Eq.(\ref{eq:condensate_corellator}) and isolating the term in the correlator containing a double insertion of the gluon field strength tensor $G_{\alpha\beta}$, corresponds to the Feynman diagram shown on the left of Fig.\ref{fig:Condensates_Non-vanishing}. Applying the partial derivatives,
\begin{equation}
\begin{split}
    \left. \frac{\partial S^{(0)}(-k-p)}{\partial p_{\rho}}\right |_{p=0} &=  \left. \frac{\partial}{\partial p_{\rho}}\frac{-\slashed{k}-\slashed{p}}{(k+p)^2}\right |_{p=0}\\
    &=\frac{2k^{\rho}\slashed{k}}{k^4}-\frac{\gamma^{\rho}}{k^2}
    \end{split}
\end{equation}
and using the relation,
\begin{equation}
\begin{split}
    G^b_{\rho \alpha}(0)\,G^c_{\sigma \beta}(0) &=\delta^{bc}\,\frac{(g_{\rho\sigma}g_{\alpha\beta}-g_{\rho\beta}g_{\alpha\sigma})}{d(d-1)(N_c^2-1)}\,G^d_{\mu \nu}(0) G^d_{\mu \nu}(0)
    \end{split}
\end{equation}
the calculation is then straight forward. After taking the trace we used FIRE \cite{Smirnov:2014hma} and ran an IBP reduction. The latter step is not necessary but it does provide us with the compact result,
\begin{equation}
K^{\braket{GG}}_{\tilde{Q}_P}(\omega_1,\omega_2)=    -\frac{4(-3+d)^2(2-3d+d^2)}{\omega_1 \omega_2}I(\omega_1)I(\omega_2)\frac{1}{4N_c\,d(d-1)}\braket{\frac{\alpha_s}{\pi}GG}
\end{equation}
where $I(\omega)$ is a 1-loop HQET integral expressible in terms of Gamma functions,
\begin{equation}
    I(\omega)=\frac{i}{(4 \pi)^{d/2}}\, (2 \omega)^{1-2 \epsilon} \Gamma(1 - \epsilon) \Gamma(2 \epsilon - 1)
\end{equation}
In order to use this in our calculation of the bag parameters, we then take the double discontinuity of the correlator and arrive at,
\begin{equation}
\begin{split}
    \rho^{\braket{GG}}_{\tilde{Q}_P} &= (e^{4\pi i \epsilon}+e^{-4\pi i \epsilon}-2)\frac{K^{\braket{GG}}_{\tilde{Q}_P}(\omega_1,\omega_2)}{(2\pi i)^2}\\
    &= \frac{\braket{\frac{\alpha_s}{\pi}GG}}{384\pi^2}
\end{split}
\end{equation}

\section{On the logarithmic divergence at $x=1$\label{sec:log_divergence}}

\begin{figure}
    \centering
    \includegraphics[width = \textwidth]{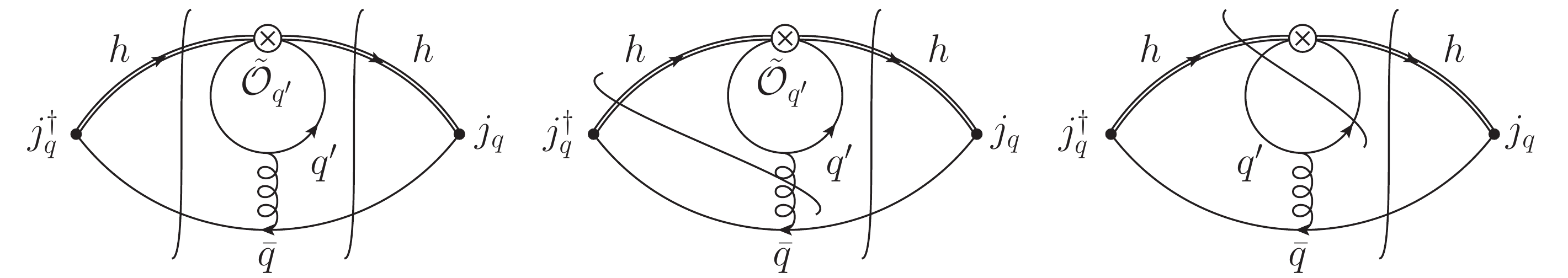}
    \caption{Cuts which yield contributions to the double discontinuity. Symmetric diagrams are not shown.}
    \label{fig:Eye_Cuts}
\end{figure}

To investigate the origin of the logarithmic divergences in the results \eqref{eq:Lifetimes_t2} for the eye contractions, we study the cuts of the relevant diagram which are contributing to the double discontinuity (see Figure~\ref{fig:Eye_Cuts}). To simplify the discussion in this appendix we only consider the scalar diagram and only work to the first order in $\epsilon$ where such a logarithm appears, but we retain the full strange-quark mass dependence in the penguin loop. Our results for the cuts (assuming $\omega_2>\omega_1$, denoted by $S_l$, $S_m$ and $S_r$ in this order for the three diagrams) are 
\begin{align}
    S_l ={}& \prod_{j=1}^3 \left(\int\frac{d^dk_j}{i\pi^{d/2}}\right)
             \frac{(-2\pi i)^4\delta(2\omega_1-2v\cdot k_1)\delta(2\omega_2-2v\cdot k_2)\delta_+(k_1^2)\delta_+(k_2^2)}{(k_1-k_2)^2[k_3^2-m_s^2][(k_3+k_2-k_1)^2-m_s^2]}\nonumber\\
        ={}& \frac{2\pi^3\Gamma(\epsilon)\Gamma(-\epsilon)}{\Gamma(1/2-\epsilon)\Gamma(1-\epsilon)\Gamma(3/2-\epsilon)\omega_1^{2\epsilon}\omega_2^{2\epsilon}m_s^{2\epsilon}} + \mathcal{O}(\epsilon^0)\,,\\
    S_m ={}& \prod_{j=1}^3 \left(\int\frac{d^dk_j}{i\pi^{d/2}}\right)
             \frac{(-2\pi i)^4\delta(2\omega_1-2v\cdot k_1)\delta(2\omega_2-2v\cdot (k_1+k_2))\delta_+(k_1^2)\delta_+(k_2^2)}{(k_1+k_2)^2[k_3^2-m_s^2][(k_3+k_2)^2-m_s^2]}\nonumber\\
        ={}& -\frac{2\pi^3\Gamma(\epsilon)\Gamma(-\epsilon)}{\Gamma(1/2-\epsilon)\Gamma(1-\epsilon)\Gamma(3/2-\epsilon)\omega_1^{2\epsilon}(\omega_2-\omega_1)^{2\epsilon}m_s^{2\epsilon}}\,,\\
    S_r ={}& \prod_{j=1}^3 \left(\int\frac{d^dk_j}{i\pi^{d/2}}\right)
             \frac{(-2\pi i)^5\delta(2\omega_1-2v\cdot k_1)\delta(2\omega_2-2v\cdot (k_1+k_2))}{k_2^2(k_1+k_2)^2}\nonumber\\
           & \times \delta_+(k_1^2)\delta_+(k_3^2-m_s^2)\delta_+((k_2+k_3)^2-m_s^2) \nonumber\\
        ={}& \mathcal{O}(\epsilon^0)\,.
\end{align}
Summing up these contributions, we find at the first non-vanishing order 
\begin{equation}
    \left.S_l + S_m + S_r\right|_{\omega_2>\omega_1} = -\frac{8\pi^2}{\epsilon} \ln\left(1-\frac{\omega_1}{\omega_2}\right) + \mathcal{O}(\epsilon^0)\,,
\end{equation}
which diverges logarithmically as $\omega_1\to\omega_2$. We reproduced this result by using our setup described in Section~\ref{subsec:method} to first compute the scalar diagram and then taking its double discontinuity. To understand this behaviour, we first note that the external momentum $p_2-p_1$ at the four-quark operator is assumed to be light-like and thus vanishes when $\omega_1=\omega_2$. Thus, in this limit the process between the two cuts in the diagram in the middle of Figure~\ref{fig:Eye_Cuts} therefore reduces to the amplitude with two external eikonal lines and one massless line which are all on-shell and is not kinematically allowed. On the other hand the processes between the two cuts of the other diagrams reduce to amplitudes with four external on-shell legs, which are kinematically possible. We further note that both the left and middle diagrams contain collinear divergences which cancel between the leading poles of both contributions, but generate the logarithms at sub-leading orders. Examining the diagrams in the 'tree' contributions, we find that there are no double-cuts which yield processes that are kinematically forbidden in the limit $\omega_1\to\omega_2$, which explains why the logarithmic divergences are only found in the 'eye' contributions. This behaviour is reminiscent of large threshold logarithms that e.g. arise in Higgs production, where infrared $1/\epsilon$ poles cancel in the sum of real and virtual corrections, but large logarithms appear because the real corrections are phase-space suppressed near the threshold. Interestingly though, the logarithms we observe here appear to be of collinear rather than soft origin.

\section{$F_q$ and $\overline{\Lambda}_q$ analysis}
\label{sec:FandLambda_SR}
For the discontinuity $\rho_\Pi(\omega)$ needed to form the sum rule of the HQET decay constant, we use the NLO result computed in
Ref.~\cite{Broadhurst:1991fc} along with the $m_s$ expanded result computed in
Ref.~\cite{King:2019lal},

\begin{eqnarray}
\rho_\Pi(\omega) & \equiv & \frac{\Pi(\omega+i0)-\Pi(\omega-i0)}{2\pi i} \\
&    =   & \frac{N_c\omega^2}{2\pi^2}\,\theta(\omega-m_s)\Bigg\{1 + \frac{m_s}{\omega} - \frac12\,\left(\frac{m_s}{\omega}\right)^2 + \dots \nonumber\\
& & + \frac{\alpha_sC_F}{4\pi}\,\Bigg[17 + \frac{4\pi^2}{3} + 3\ln\frac{\mu_\rho^2}{4\omega^2} +\left(20 + \frac{4\pi^2}{3} + 6\ln\frac{\mu_\rho^2}{4\omega^2} - 3\ln\frac{\mu_\rho^2}{m_s^2}\right)\,\frac{m_s}{\omega} \nonumber\\
& & + \left(1 - \frac92 \ln\frac{\mu_\rho^2}{4\omega^2} + 3\ln\frac{\mu_\rho^2}{m_s^2}\right)\,\left(\frac{m_s}{\omega}\right)^2+ \dots\Bigg] + \mathcal{O}(\alpha_s^2)\Bigg\} \\ \nonumber
& & - \frac{\langle\bar{s}s\rangle}{2}\delta(\omega)\left[1+6\frac{\alpha_sC_F}{4\pi} + \mathcal{O}(\alpha_s^2)\right]+\frac{\langle\bar{s}i\sigma_{\mu\nu}G^{\mu\nu}s\rangle}{32} \delta''(\omega)\left[1 + \mathcal{O}(\alpha_s)\right] + \mathcal{O}(\Lambda^6).
\label{eq:rho_Pi_ms_exp}
\end{eqnarray}

After plugging Eq.(\ref{eq:rho_Pi_ms_exp}) into Eq.(\ref{eq:DecayConstantSR}), logarithmic terms of the form $\log(\mu^2/m_s^2)$ can be resummed by switching to the $\overline{\text{MS}}$ scheme,

\begin{equation}
m_s = \bar{m}_s(\mu_\rho) \left[1 + \frac{\alpha_s(\mu_\rho)C_F}{4\pi}\left(4+3\log\left(\frac{\mu_\rho^2}{\bar{m}_s^2(\mu_\rho)}\right)\right) + \dots\right]\,.
\end{equation}
We also note that $m_s$ terms arising from the lower integration cut in Eq.(\ref{eq:DecayConstantSR}) were not expanded in $m_s$.\\

The running of the quark condensates takes the form 
[see Ref.~\cite{Bagan:1991sg}]

\begin{align}
\langle\bar{s}s\rangle(\mu_\rho) & = \langle\bar{s}s\rangle(\mu_0)\left[\frac{\alpha_s(\mu_\rho)}{\alpha_s(\mu_0)}\right]^{\frac{\gamma_0^{(3)}}{2\beta_0}}
\times\left[1+\frac{\alpha_s(\mu_\rho)-\alpha_s(\mu_0)}{4\pi}\frac{\gamma_0^{(3)}}{2\beta_0}\left(\frac{\gamma_1^{(3)}}{\gamma_0^{(3)}}-\frac{\beta_1}{\beta_0}\right)\right] \,, \nonumber\\
\langle\bar{s}i\sigma_{\mu\nu}G^{\mu\nu}s\rangle(\mu_\rho) & = \langle\bar{s}i\sigma_{\mu\nu}G^{\mu\nu}s\rangle(\mu_0)\left[\frac{\alpha_s(\mu_\rho)}{\alpha_s(\mu_0)}\right]^{\frac{\gamma_0^{(5)}}{2\beta_0}} \,,
\end{align}

with $\gamma_0^{(3)}=-8$, $\gamma_0^{(5)}=-4/3$, $\gamma_1^{(3)}=-404/3+40n_f/9$, $\beta_0=11-2n_f/3$ and $\beta_1=102-38n_f/3$.
The logarithmic derivative of Eq.(\ref{eq:DecayConstantSR}) furthermore gives us a sum rule for the mass difference $\overline{\Lambda}_s$ in the form 
[see Ref.~\cite{Bagan:1991sg}]

\begin{equation}
\overline{\Lambda} = t^2 \frac{\frac{d}{dt}\int\limits_0^{\omega_c}d\omega\, e^{-\frac{\omega}{t}}\rho_\Pi(\omega)}{\int\limits_0^{\omega_c}d\omega\, e^{-\frac{\omega}{t}}\rho_\Pi(\omega)}
= \frac{\int\limits_0^{\omega_c}d\omega\, \omega\,e^{-\frac{\omega}{t}}\rho_\Pi(\omega)}{\int\limits_0^{\omega_c}d\omega\,e^{-\frac{\omega}{t}}\rho_\Pi(\omega)} \, .
\end{equation}

\begin{figure}
     \centering
     \begin{subfigure}[b]{\textwidth}
         \centering
         \includegraphics[width=\textwidth]{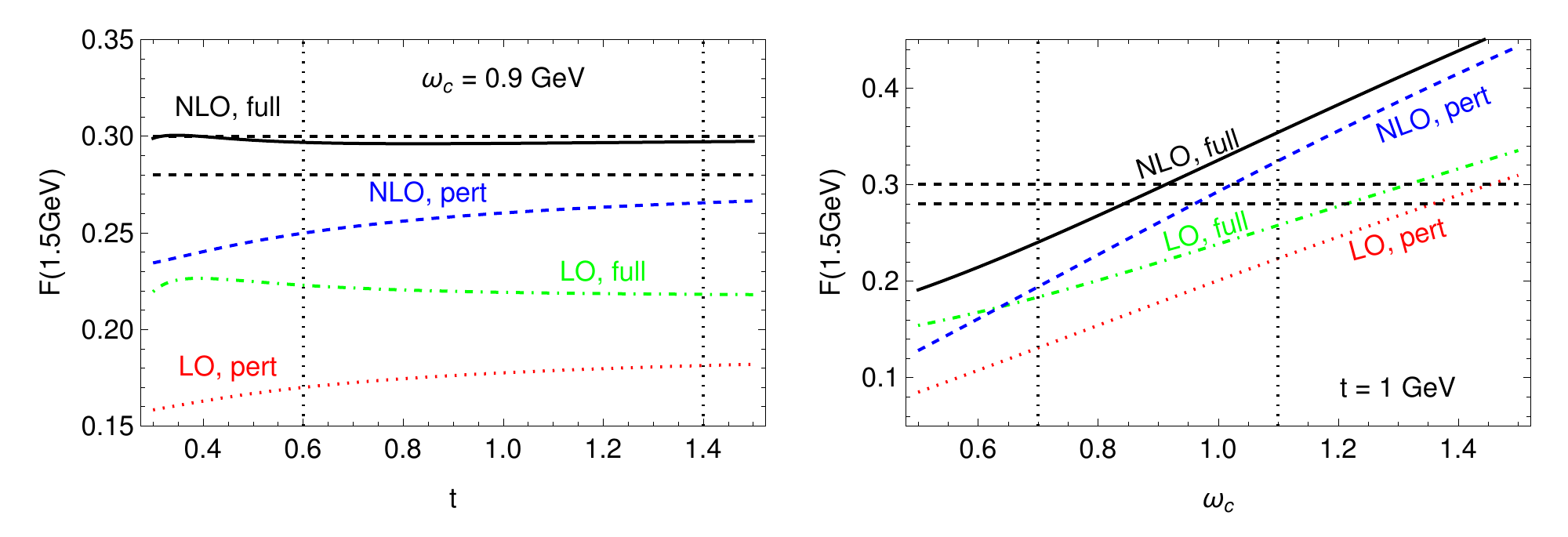}
     \end{subfigure}
     \vspace{0.5mm}
     \begin{subfigure}[b]{\textwidth}
         \centering
         \includegraphics[width=\textwidth]{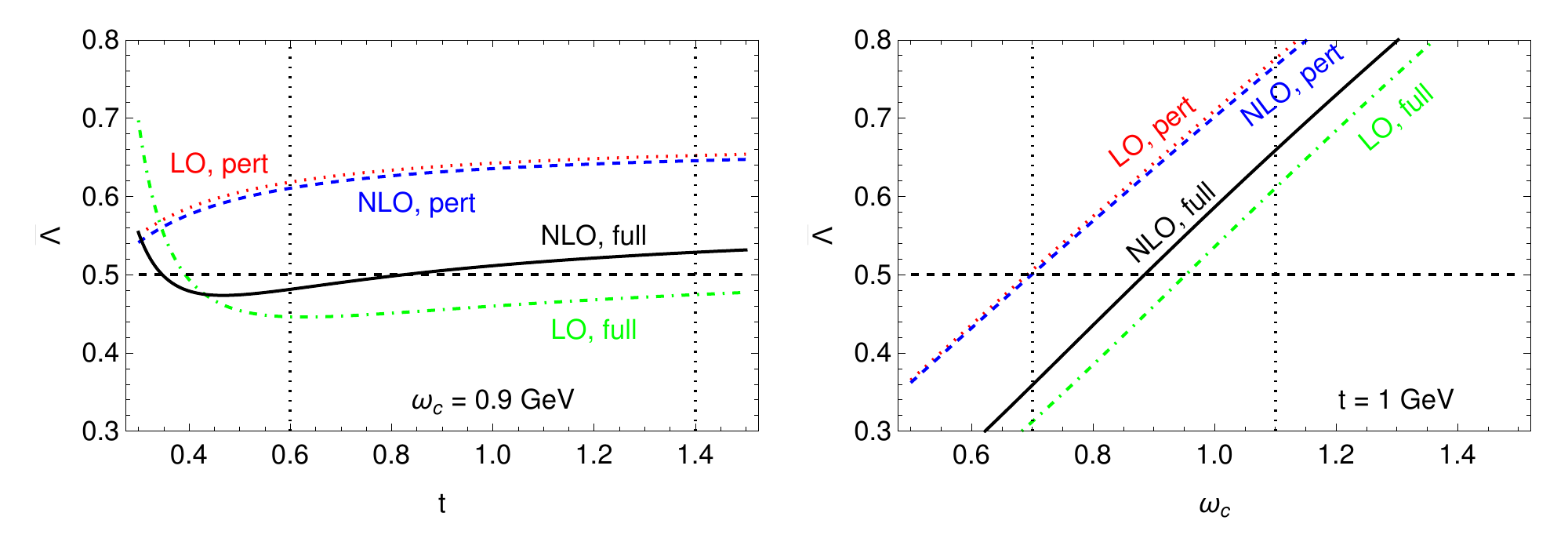}
     \end{subfigure}
     \caption{Dependence of the sum rule results for $F(\mu)$ (top) and $\overline{\Lambda}$ (bottom) on the Borel parameter $t$ (left) and the continuum cutoff $\omega_c$ (right) in the $B_q$ system. }
     \label{fig:F_Lambdabar}
\end{figure}
\begin{figure}
  \centering
     \begin{subfigure}[b]{\textwidth}
         \centering
         \includegraphics[width=\textwidth]{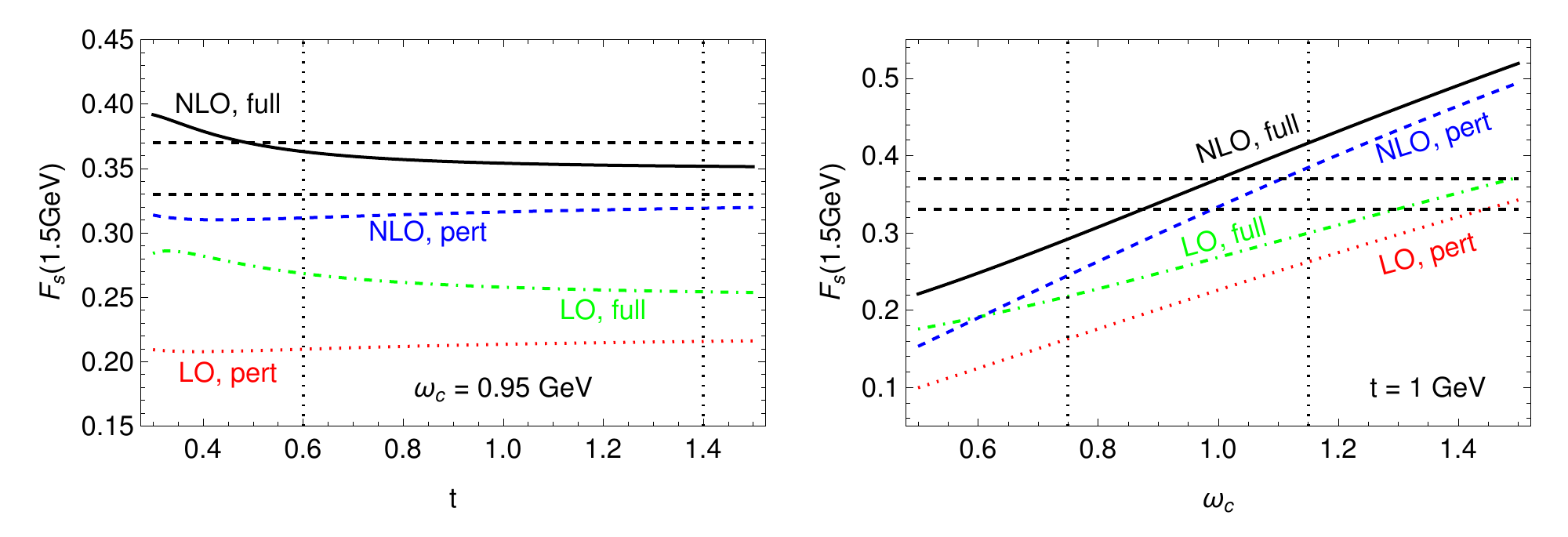}
     \end{subfigure}
     \vspace{0.5mm}
     \begin{subfigure}[b]{\textwidth}
         \centering
         \includegraphics[width=\textwidth]{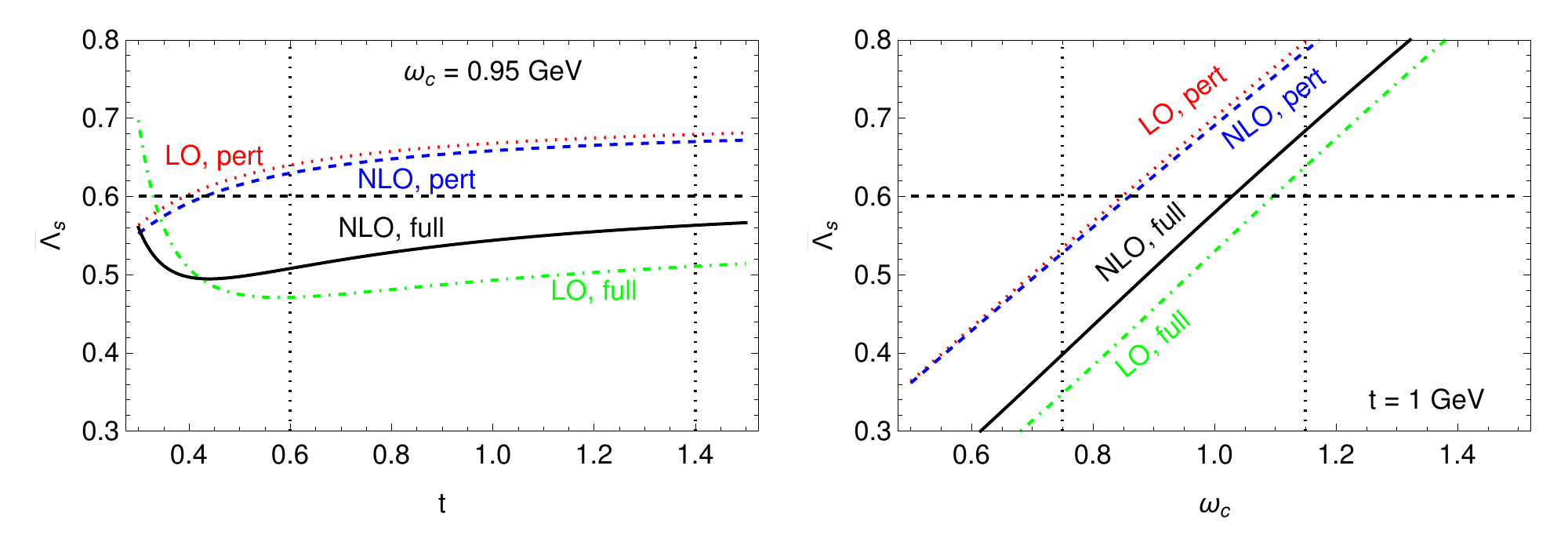}
     \end{subfigure}
     \caption{Dependence of the sum rule results for $F_s(\mu)$ (top) and $\overline{\Lambda}_s$ (bottom) on the Borel parameter $t$ (left) and the continuum cutoff $\omega_c$ (right) in the $B_s$ system.}
     \label{fig:Fs_Lambdabars}
\end{figure}

To determine the appropriate ranges for the Borel parameters $t_i$ and the continuum cutoff $\omega_c$ in our bag parameter analysis, we consider the sum rules for the meson-heavy quark mass difference $\overline{\Lambda}_q$ and the HQET decay constant $F$ and compare with the values found in the literature. The values of the HQET decay constants,

\begin{equation}
F(1.5\,\text{GeV}) = (0.29\pm0.01)\,\text{GeV}\,,\hspace{1cm}F_s(1.5\,\text{GeV}) = (0.35\pm0.02)\,\text{GeV}\,,
\end{equation}

are determined from the static results of the ALPHA collaboration from
Ref.~\cite{ALPHA:2014lwy} by matching at the scale $\bar{m}_b(\bar{m}_b)$ and evolving the HQET decay constants down to the scale 1.5GeV. For the mass differences we use, $\overline{\Lambda}_q=0.5$ and $\overline{\Lambda}_s=0.6$ for the $B_q$ and $B_s$ mesons respectively. We find the behaviour shown in Figures \ref{fig:F_Lambdabar} and \ref{fig:Fs_Lambdabars} from which we determine the following intervals:

\begin{align}
B_q:&\hspace{1cm}t=(1.0\pm0.4)\,\text{GeV}\,,\hspace{0.5cm}\omega_c=(0.90\pm0.2)\,\text{GeV}\,,\nonumber\\
B_s:&\hspace{1cm}t=(1.0\pm0.4)\,\text{GeV}\,,\hspace{0.5cm}\omega_c=(0.95\pm0.2)\,\text{GeV}\,.
\label{eq:SRparameters}
\end{align}

\bibliographystyle{JHEP-cerdeno}
\bibliography{refs}
\end{document}